%% file: main.tex
\PassOptionsToPackage{dvipsnames}{xcolor}
\documentclass[acmsmall]{acmart}

\AtBeginDocument{%
  \providecommand\BibTeX{{%
    \normalfont B\kern-0.5em{\scshape i\kern-0.25em b}\kern-0.8em\TeX}}}

\setcopyright{acmcopyright}
\copyrightyear{2024}
\acmYear{2024}
\acmDOI{XXXXXXX.XXXXXXX}

\acmJournal{TOSEM}
\acmVolume{1}
\acmNumber{1}
\acmArticle{1}
\acmMonth{10}

\usepackage{tikz}

\usepackage{multirow}
\usepackage{blindtext}
\usepackage{hyperref}
\usepackage{hyperxmp}
\usepackage{listings}
\usepackage{tcolorbox}
\usepackage{subcaption}
\usepackage{tikz}
\definecolor{darkgreen}{rgb}{0.0, 0.5, 0.0}
\newcommand{\codebert}{\text{CodeBERT}}
\newcommand{\codellama}{\text{CodeLlama-7B-HF}}
\newcommand{\starcoder}{\text{StarCoder2-7B}}
\newcommand{\qwen}{\text{Qwen2.5-Coder-7B}}
\newcommand{\ct}{\text{CodeT5+}}
\usepackage{svg}

\begin{document}

%%\textcolor{blue}{TEXT IS HERE}
\title{MANTRA: a Framework for Multi-stage Adaptive Noise TReAtment During Training}

%%Script to Smart: How Language Models Embrace R Code
%%Can AI Models 'Speak' R? Assessing Pre-trained Models' Performance in R Landscapes
%%Decoding R: How Pre-trained AI Models Navigate the R Language

\author{Zixiao Zhao}
\email{zixiaosh@student.ubc.ca}
\orcid{0009-0008-6947-2723}
\affiliation{%
  \institution{University of British Columbia}
  \city{Kelowna}
  \state{BC}
  \country{Canada}
}

\author{Fatemeh H. Fard}
\email{fatemeh.fard@ubc.ca}
\orcid{1234-5678-9012}
\affiliation{%
  \institution{University of British Columbia}
  \city{Kelowna}
  \country{Canada}
}

\author{Jie JW Wu}
\email{jie.jw.wu@mtu.edu}
\orcid{0000-0002-7895-2023}
\affiliation{%
  %\institution{University of British Columbia}
  %\city{Kelowna}
  %\country{Canada}
  \institution{Michigan Technological University}
  \city{Houghton}
  \state{Michigan}
  \country{United States}
}

\renewcommand{\shortauthors}{Zhao et al.}

\begin{abstract}
The reliable application of deep learning models to software engineering tasks 
% such as code summarization, bug detection, and code generation, 
hinges on high-quality training data. Yet, large-scale repositories inevitably introduce noisy or mislabeled examples that degrade both accuracy and robustness. 
While Noise Label Learning (NLL) has been extensively studied in other fields, there are a few works that investigate NLL in Software Engineering (SE) and Large Language Models (LLMs) for SE tasks. 
In this work, we propose \textbf{MANTRA}, a \textbf{M}ulti-stage \textbf{A}daptive \textbf{N}oise \textbf{TR}e\textbf{A}tment framework that embeds noise diagnosis and mitigation directly into the fine-tuning process of code-Pretrained Language Models (PTM) and code-LLMs.

We first investigate the effect of noise at varying levels on convergence and loss trajectories of the models.
Then we apply an adaptive dropout strategy guided by per-sample loss dynamics and Gaussian Mixture Model clustering to exclude persistently noisy points while preserving clean data. 
Applying to code summarization and commit intent classification, our experiments reveal that some LLMs are more sensitive to noise than others. However, with MANTRA, the performance of all models in both tasks is improved. 
MANTRA enables researchers and practitioners to reduce the impact of errors introduced by the dataset in training, saves time in data cleaning and processing, while maximizing the effect of fine-tuning.
% , thereby uncovering how different architectures respond to mislabeled or corrupted samples. Our experiments reveal that both PTM and LLMs are affected to varying degrees when confronted with noisy training data, and this effect is reflected in their training process. Building on these insights, we propose an adaptive dropout strategy guided by per-sample loss dynamics and Gaussian Mixture Model clustering to exclude persistently noisy points while preserving informative diversity selectively. Using our approach, we can achieve a 10\% improvement on the original dataset (using Qwen2.5-coder for code summarization), and for the noisy training set, we can obtain a 5\% to 10\% improvement. MANTRA enables researchers and practitioners to reduce the impact of errors introduced by the dataset in training, saves time in data cleaning and processing, while maximizing the effect of fine-tuning.
\end{abstract}




\received{01 October 2024}
\received[revised]{2024}
\received[accepted]{10 April 2024}

\maketitle

\input{sections/intro}
\input{sections/relatework}
\input{sections/methodology}
\input{sections/result}
\input{sections/discussion}
\input{sections/Threats}
\input{sections/conclusion}

\bibliographystyle{ACM-Reference-Format}
{\footnotesize\bibliography{references}}

\end{document}

%% file: sections/intro.tex
\section{Introduction}
The swift progress of Artificial Intelligence (AI) in Software Engineering (SE) has driven the adoption of deep learning models across multiple SE tasks~\cite{comparison2023, Ahmed, codellama, GPT2, GPT3}. These models, which include Code-Pretrained Language Models (PTM) and Code-Large Language Models (LLMs), are utilized for different software engineering tasks like code summarization and bug detection~\cite{CodeXGLUE}.
The effectiveness of these models greatly depends on the quality of the training data, since noisy data can cause incorrect predictions, poorer generalization, and unreliable results~\cite{LaxmiNarayanaChejarla_2025, 10.1145/3728957, 11052608, shah2024towards, qiao-etal-2022-selfmix-GMM}. 
Studies have shown that smaller models can achieve close or even better performance than larger models when trained on a higher-quality dataset~\cite{textbooks}.
However, obtaining high-quality, noise-free datasets for code-related tasks remains an ongoing challenge~\cite{codet5+, 11025700, li2022robust-GMM}, especially as these datasets are frequently derived from large-scale platforms like GitHub and StackOverflow, where manual verification is infeasible due to data volume and complexity~\cite{li2022robust-GMM}.

% Noise in training data often manifests itself as label noise, where samples do not correspond accurately to their target labels, creating significant obstacles in supervised learning tasks~\cite{Hallucination}. For instance, in code summarization, descriptions that fail to match paired code snippets introduce noise~\cite{Hallucination}; in bug detection, fully functional code mislabeled as buggy can misguide the model~\cite{10.1145/3746225}. 
Given the inherent complexity of program data, noisy data exists across different benchmarks~\cite{shah2024towards}, such as when descriptions fail to match paired code snippets for code summarization~\cite{Hallucination} or a fully functional code is mislabeled as buggy~\cite{10.1145/3746225}.
The presence of noise in the training data can cause slow convergence and gradient instability~\cite{shah2024towards}, affecting models' robustness in real-world applications~\cite{LEVER2025102523, 9524547}. 
% the datasets can exhibit a proportion of such noisy and complex labels, which degrades model accuracy and, importantly, reduces the model's robustness in real-world applications~\cite{LEVER2025102523, 9524547}. 
Robustness in this context is defined as the ability of models to maintain performance when confronted with noisy data. 
% \JW{should we emphasize the issue somewhere that in literature we don't know the relationship of noise vs loss/convergence of LLMs (addressed by RQ1)?}

To tackle these challenges, many studies have concentrated on Noise Label Learning (NLL) to create methods for handling noise in datasets~\cite{Zhu2021DetectingCL, AdversarialPerturbations,Jiang2022AnIF, Kim2024LearningWN}, with the goal of improving model robustness by identifying and mitigating the impact of erroneous labels~\cite{Zhu2021DetectingCL, li2022robust-GMM}. 
However, NLL studies in SE are limited in number and scope. They are mostly empirical studies or applied on classification tasks~\cite{wang2024empirical, 9796240, Dau_2022, li2022robust-GMM, shah2024towards}, with no NLL approach in SE that works for a variety of tasks like classification and generative tasks. Additionally, these methods tend to be optimized for small to medium-sized models, leaving a gap in effective noise management techniques for LLMs in software engineering. 
On the other hand, the NLL research in other fields typically rely on techniques that require pre-calculated confidence estimates~\cite{Kye2021LearningWN}, or dedicated or task-specific models for noise handling~\cite{Co-teaching, EarlyStoppingforLearning}. 
While some researchers have proposed self-training techniques to handle noisy data during training~\cite{Yuan_2024, li2022robust-GMM}, to the best of our knowledge, no work in SE has addressed self-training for different task types in various-sized models, including PTMs and LLMs.

Our research seeks to fill these gaps by proposing a generalized framework for managing noisy data in code-related tasks, focusing on PTMs and LLMs to improve model robustness in the presence of noisy data. 
% We consider robustness in this context as the ability of models to maintain performance when confronted with noisy data. 
In this work, we present \textbf{MANTRA}, a \textit{\textbf{M}ulti-stage \textbf{A}daptive \textbf{N}oise \textbf{TR}e\textbf{A}tment} designed specifically to tackle the issues of noisy labels and data anomalies in tasks related to SE. 
This study comprises two core stages:

\begin{itemize}
    \item \textbf{Investigation of Noise Effects on Model Dynamics:} We conduct a systematic investigation into how noise influences model convergence and loss trajectories during training, introducing synthetic noise at different levels to simulate real-world data inconsistencies. This contribution provides novel insights into how noisy data points affect model stability.
    
    \item \textbf{Dynamic Noise Management via Adaptive Dropout Strategies:} We develop a dynamic method for identifying and dropping noise-prone data points during fine-tuning. This approach combines adaptive dropout with noise-resistant regularization techniques, offering a targeted solution for mitigating noise effects without compromising data diversity. This contribution is particularly valuable for fine-tuning in real-world applications, where clean datasets are rare.
    
\end{itemize}

We experiment with five different code models, both from PTM and LLMs: 
CodeBERT, CodeT5+, CodeLlama-7B-HF, StarCoder2-7B, and 
Qwen2.5-Coder-7B.  
The models are assessed on two SE tasks:  
(i) \textit{code summarization}, and 
(ii) \textit{commit intent classification}.
Each model is evaluated in four label noise settings for each task: 0\%, 5\%, 10\%, and 15\%, with and without MANTRA. 
Our findings support that PTMs have different loss distributions for clean and noisy data~\cite{li2022robust-GMM}; however, this finding depends on the studied LLMs. CodeLlama shows different distributions for clean and noise samples, but, this is less pronounced for the other two LLMs. 
Additionally, some LLMs are more robust to noisy data than others. In all models, both PTMs and LLMs, MANTRA makes the models robust, with minimal performance degradation in the presence of label noise. 
% First, we characterise the \emph{per-sample loss behaviour} of all five models: clean examples exhibit smoothly decaying losses, whereas noisy ones oscillate with high variance and make the model hard to converge.  
% Second, we embed our proposed \textsc{MANTRA} module into the LoRA fine-tuning loop, allowing the model to downweight samples whose loss trajectories match the noisy pattern during fine-tuning. 
% \FHF{Is this result observed for all tasks? You should explicitly mention the two gains for each task. Also, you can use words like "up to 10\%" if this is not always the case. I think results summary can be written better, conveying the benefits of your work.}
% , and achieve
% a \shawn{up to }10\% improvement in performance on the original dataset and obtain a 5\% to 10\% improvement in performance using the noise dataset. 

% Specifically, a robust model should exhibit minimal performance degradation in the presence of label noise, effectively adapt to uncertain inputs.

% Our approach enables generalization to different tasks, without requiring calibrations to accommodate task needs, as it directly works with the loss scores of the models. Additionally, as our work is task agnostic, and is applied during training, it reduces the need to curate a dataset with labeled noisy and clean samples and thus the reliance on a separate noise detection model. 

MANTRA offers a task‑agnostic way to spot and correct label noise as training unfolds, strengthening the reliability and robustness of large code models without per‑task retuning. Restraining noise at each stage of learning yields more accurate and generalizable systems and sets the groundwork for future research on noise‑resilient training in SE applications.

The primary contributions of this research are: First, by analyzing loss trajectories across two code intelligence tasks for PTMs and LLMs, we shed new light on how label noise behaves.  
Second, we introduce \textsc{MANTRA}, a model and task-agnostic framework that detects and down-weights suspected noisy instances \textit{during} fine-tuning. 
Finally, our experiments demonstrate that \textsc{MANTRA} consistently strengthens all five models in code summarization and commit-intent classification. We also provide a replication package for our study: \url{https://anonymous.4open.science/r/MANTRA-3653/}.

%% file: sections/relatework.tex
\section{Related Works}

% \FHF{Look at the 14 citations of this work ( Robust Learning of Deep Predictive Models from Noisy and Imbalanced Software Engineering Datasets), citation 26 in your paper. There are some relevant papers that you need to bring here and mention explicitly the difference of your work with them. Soem are related to noise data and learning with noise in Dl and LLMs for code. Also take a look at the citations of those papers and see if you can find sth interesting that we need to address. You can also get inspiration and learn how to write some parts of your paper.}

Studies in software engineering and AI fields have investigated issues concerning data quality~\cite{LaxmiNarayanaChejarla_2025, 10.1145/3728957, 11052608} and their effects on model performance \cite{Tantithamthavorn2024}, and investigating approaches to ensure high quality data~\cite{10.1145/3733599}. 
% Research has identified various data quality antipatterns specific to software analytics, revealing their significant effects on software defect prediction models \cite{Tantithamthavorn2024}. 
% Similarly, other works highlight the importance of data de-duplication to enhance quality in computational social science research \cite{Mu2024}.
% ~\citet{10.1145/3733599} investigate collaborative fine-tuning methods to ensure high data quality in LLMs without compromising data privacy.
% Given the importance of high-quality data for model training, various strategies for learning from noisy labels have been proposed, falling primarily into three categories: Filtering Strategies, Noise Estimation  Methods, and Training Strategies.
% In the following, we review the related works into three categories: Filtering Strategies, Noise Estimation  Methods, and Training Strategies.
% by evaluating prediction confidence as an indicator of label reliability. 
% \FHF{Based on this first sentence, why do you say in the intro, that current appraoches need a separate model for training for noise and a separate model for training for the task? This first sentence implies that your work falls in the filtering strategy category. If so, we should change parts of intro and explicility mention the differences and advantages of your work compared to works in this category.}
% \paragraph{Filtering Strategies}
% Filtering strategies 
Sample‑selection methods filter out noise by estimating noise rates and confidence thresholds~\cite{Confidentlearning}, calculating network-based p-values framed as a multiple-hypothesis testing problem~\cite{Yu2023DelvingIN}, 
applying unsupervised methods such as K-Nearest Neighbor to enhance the model’s ability to detect noise~\cite{Zhu2021DetectingCL}, using methods like causal analysis~\cite{AdversarialPerturbations} and co-training~\cite{Li2020DivideMixLW}, or applying thresholds to separate noise and clean data~\cite{qiao-etal-2022-selfmix-GMM}.

% \citet{Confidentlearning} introduced an approach that estimates the proportion of noisy and correct labels using cross-validation results, incorporating thresholds and covariance matrices to filter out samples with confidence scores that are highly inconsistent with their labels. Confidence regularization techniques have also been employed to control model overfitting by guiding the model towards a more global and balanced understanding of sample confidence~\cite{instance-dependent}. In a different approach, detecting noisy labels in nominally clean datasets has been framed as a multiple-hypothesis testing problem~\cite{Yu2023DelvingIN}, where neural network-based p-values are calculated to assess sample suitability for training. 

% From a model-fitting perspective, unsupervised methods such as K-Nearest Neighbor (KNN) aggregation help characterize sample clusters, enhancing the model’s ability to detect noise, while feature quality analysis aids in screening ambiguous samples~\cite{Zhu2021DetectingCL}. Semi-supervised approaches target model generalization, using methods like causal analysis~\cite{AdversarialPerturbations} and co-training~\cite{Li2020DivideMixLW} to improve the model's capacity to identify and separate noisy samples from clean data.

% \paragraph{Noise Estimation Methods}
Other approaches~\cite{Jiang2022AnIF, Kye2021LearningWN} focus on estimating noise generation matrices to minimize the noise generation process, or provide solvable conditions for noisy data in regression~\cite{Englesson2023LogisticNormalLF}. 
Other researchers employed Bayesian methods to improve estimation accuracy~\cite{Yao2023WhichIB} or proposed new loss functions to improve robustness during training~\cite {Peerloss}.

~\citet{earlylearning, EarlyStoppingforLearning} strengthen noise robustness by refining loss functions and employing early stopping to prevent overfitting to noise. Co-teaching involves one model identifying noise patterns in the labels and giving feedback to the other model, helping it better distinguish between noisy and clean samples~\cite{Co-teaching, Li2020DivideMixLW}.~\citet{Yuan_2024} explored collaborative frameworks where small models interact with LLMs to identify and relabel noisy samples dynamically during training.~\citet{Ye2025CalibratingPL}~explore refinement of LLM-generated labels during training to calibrate pre-trained classifiers against label noise. ~\citet{Kim2024LearningWN} has used two EM procedures, enabling optimization under label noise without requiring external supervision. Hyperspherical margin weighting to adaptively reweight samples based on their angular distances in feature space is proposed~\cite{Zhang2024LearningWN}.
~\citet{luo2024} use a multi-expert system, context-enhanced denoising, and entropy-based sample selection to detect and relabel noisy responses. ~\citet{wang2025}~introduces a framework that injects noise into a poisoning expert module during training to improve the robustness.

The studies on noise label learning in SE, however, is limited compared to the NLP and computer vision fields. An empirical study on three SE tasks, program classification, vulnerability detection, and code summarization, is conducted, comparing NLL approaches on smaller and larger models on different noise types~\cite{wang2024empirical}. Another study uses the semantic difference between samples to detect noise for the security defect task~\cite{9796240}. 
Data influence methods, Influence Function, and TracIn, are applied to detect and remove noisy samples in the source code corpus for two classification tasks~\cite{Dau_2022}.
% ~detect and remove noisy samples in the source code corpus by introducing data influence methods (Influence Function and TracIn), thus improving the performance and robustness of the neural source code model in classification tasks. 
A two-stage robust training framework is proposed in~\cite{li2022robust-GMM} to solve the problem of mislabeling and class imbalance in code-based classification datasets.
% \cite{li2022robust-GMM}~proposed a two-stage robust training framework to effectively solve the problem of the coexistence of mislabeling and category imbalance in software engineering data by combining feature representation learning and classifier retraining, so as to improve the performance of deep prediction models.
Finally, a recent empirical study investigated data bugs across different data modalities, revealing their impact on training dynamics, thereby offering actionable insights for data cleaning and model monitoring~\cite{shah2024towards}.

\textbf{Differences of our work.}
The closest studies to our approach are~\cite{qiao-etal-2022-selfmix-GMM, li2022robust-GMM}. The former~\cite{qiao-etal-2022-selfmix-GMM} tailors GMM for classification tasks in NLP and erases the original labels of the identified samples for semi-supervised self-training. 
The latter ~\cite{li2022robust-GMM} applies GMM on similarity scores, which need to be calculated based on ground truth or class representative for SE classification tasks. 
Furthermore, the earlier works~\cite{qiao-etal-2022-selfmix-GMM, li2022robust-GMM, wang2024empirical} only use smaller deep learning and pre-training language models (e.g., Roberta, CodeBERT, UniXCoder), without exploring code large language models. 
In contrast, MANTRA works for classification \textit{and} generative tasks, applied on PTMs and LLMs. 

%% file: sections/methodology.tex
\section{Methodology} \label{sec: methodology}

\begin{figure}[t]
  \centering
  \includegraphics[width=\columnwidth]{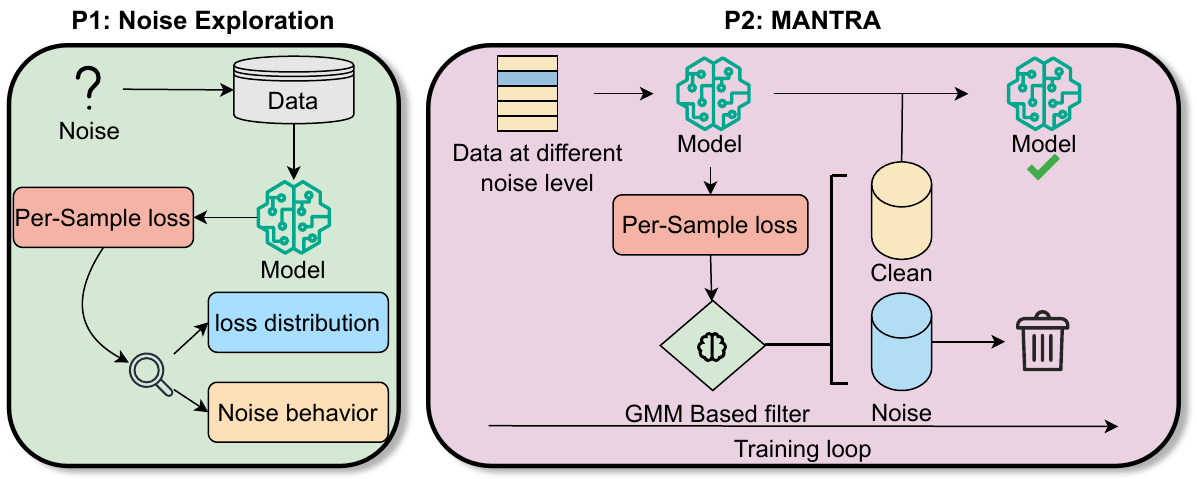}
  \caption{The workflow for this study. P1 explores loss distribution and noise behaviour, P2 applies MANTRA during training.}
  \label{fig:flow}
  \vspace{-0.5em}
\end{figure}

In this section, we will examine the research questions, the tasks and models analyzed, and our method for addressing each research question. 

\subsection{Research Questions}
We investigate the answer to the following questions. 
% We defi?ne the following research questions to guide our investigation into the impact of noisy data and the effectiveness of noise management techniques in enhancing model robustness for SE code-related task.

\textbf{RQ1: How does noise affect the loss trajectory and convergence behavior of PTMs and LLMs during training?} \\
    Understanding how noise influences training dynamics is essential for designing robust models capable of handling noisy data. 
    Noisy samples can disrupt the learning process by causing inconsistent loss patterns, slower convergence, or even overfitting to erroneous labels~\cite{Kim2024LearningWN}. Earlier studies indicate that clean and noisy samples exhibit different patterns in loss values and convergence rates~\cite{qiao-etal-2022-selfmix-GMM}. 
    This research question seeks to investigate the impact of noise on the tasks being studied, specifically for PTMs and LLMs. We introduce different levels of synthetic noise and track the loss trajectories of samples during training. 
    % Key metrics, such as loss variance, performance, and convergence rates, are analyzed to quantify the effect of noise on model behavior.

% \textbf{RQ2: How does MANTRA ’s dynamic noise detection and filtering mechanism during fine-tuning affect the generalization and robustness of large language models in software engineering code-related tasks?} \\
\textbf{RQ2: How does MANTRA affect the performance of the models in the presence of noise?}

This RQ investigates the effectiveness of MANTRA in improving robustness to noisy labels during fine-tuning. MANTRA identifies noise samples exhibiting inconsistent behavior and gradually excludes them via an adaptive dropout mechanism. This delay in exclusion ensures initial exposure to diverse data while minimizing long-term overfitting to corrupted labels. We compare the performance of the models trained both with and without MANTRA.

\subsection{Approach}
\label{sec: Approach}
Figure~\ref{fig:flow} presents our approach. The first part (RQ1) explores the loss distribution of the models, and the second part (RQ2) applies MANTRA on the studied tasks. Details of each part are below. 

% In this section, we will provide a detailed experimental setup of each of the studied RQs.

\subsubsection{RQ1:} 

To address the first research question on how noise impacts loss trajectory and convergence behavior, we begin by introducing controlled noise into the datasets of each task randomly. Noise is added to the training portion of the datasets for each task, whereas the validation and test sets are left unchanged. We apply noise at three levels of intensity, 5\%, 10\%, and 15\%, and consider 0\% noise as the clean data.
Here, 5\%, 10\%, and 15\% denote the proportion of label corruption applied only to the training split, with class priors preserved; 0\% is the clean control. This enables us to assess how varying levels of noise affect the model's performance. During training, we monitor loss trajectories across epochs for both clean and noisy samples. Observing fluctuations and inconsistencies in the loss values helps identify distinctive patterns associated with noisy data, such as slower or inconsistent loss reduction for noisy samples. Choosing 5\%, 10\%, and 15\% label corruption lets us pinpoint the noise threshold at which each model’s performance starts to break down and measure how well adaptive filtering methods recover accuracy as corruption increases. Additionally, we perform a convergence analysis by comparing the model’s rate of convergence on datasets with varying noise densities, examining how different levels of noise affect the model’s stability in reaching optimal performance. Losses are used to quantify the impact of noise. This analysis provides insights into how different ratios of noisy data influence the model’s training dynamics, laying the groundwork for adaptive noise handling strategies in subsequent phases.

\subsubsection{RQ2:} 
The second research question explores whether dynamically identifying and dropping noisy samples during fine-tuning can enhance model robustness.
% Using insights gained from the first phase, 
We establish criteria to recognize noise-prone samples based on their loss trajectories. 
During fine-tuning, we apply an adaptive dropout strategy, removing samples that show persistent noise-like behavior after a specific number of epochs. This approach balances the model’s exposure to diverse data with the need to avoid overfitting to noisy samples. By selectively removing samples only after the model has had an initial opportunity to learn from them, we enable the model to generalize more effectively. 
We then compare the performance of the model 
% in terms of validation loss
% , accuracy, and generalization gap 
versus when it was trained without dynamic dropout. 
% This comparison helps us assess whether this adaptive strategy can enhance the model’s ability to perform reliably when confronted with noisy data in real-world applications.

To identify the noisy data, we use Gaussian Mixture Models (GMM) and Bayesian Information Criterion (BIC). GMMs are employed to determine whether the dataset comprises multiple underlying clean distributions. GMMs model the data as a mixture of Gaussian components, with each component defined by its own mean, variance, and weight. The parameters of these components are estimated using the Expectation-Maximization (EM) algorithm, which iteratively maximizes the likelihood of the data under the model.

E-step: \(w_j^{(i)} = p\bigl(z^{(i)} = j \mid x^{(i)};\phi,\mu,\Sigma\bigr)\) and

M-step: \[
          \phi_j \;:=\; \frac{1}{m}\,\sum_{i=1}^{m} w_{j}^{(i)},
          \qquad
          \mu_j  \;:=\; \frac{\sum_{i=1}^{m} w_{j}^{(i)}\,x^{(i)}}{\sum_{i=1}^{m} w_{j}^{(i)}},
        \]
        \[
          \Sigma_j \;:=\;
          \frac{\sum_{i=1}^{m} w_{j}^{(i)}
                 \bigl(x^{(i)} - \mu_j\bigr)\bigl(x^{(i)} - \mu_j\bigr)^{\!\top}}
               {\sum_{i=1}^{m} w_{j}^{(i)}} .
        \]

To select the optimal number of components in the GMM, we use the Bayesian Information Criterion, a measure that balances goodness-of-fit and model complexity. BIC is defined as: \(\mathrm{BIC} = -2\ln L + k \ln n\), where \(L\) is the likelihood of the model given the data, \(k\) represents the number of parameters, and \(n\) denotes the number of data points. The BIC applies a more substantial penalty for the number of parameters, which makes it especially effective for selecting simpler models when dealing with large datasets.

The BIC values are computed for GMMs with varying numbers of components, and the model with the lowest BIC is selected as the optimal representation of the data. This ensures a balance between accurately capturing the underlying distributions and avoiding overfitting, providing a robust method to identify mixtures of normal distributions in the dataset.

\subsection{Tasks and Datasets}
We choose different tasks, including generative and classification, which represent the generalizability of \textsc{MANTRA}.

\textbf{Code Summarization} aims to generate summaries of source code~\cite{CodeBERT, GraphCodeBERT, Zhu2019AutomaticCS}, 
it has been studied in several papers~\cite{bert, codesearchnet, 10.1145/3728963, 11071936, wang2024empirical}, therefore chosen in our study. 
We use the Python subset of CodeSearchNet~\cite{codesearchnet} dataset for our work.
To make synthesized noise, replace the sample's code summary with randomly selected tokens, preserving the length of the summary unchanged. For fine-tuning LLMs on code summarization, we randomly selected 10,000 samples from the CodeSearchNet~\cite{codesearchnet} Python dataset due to resource limitations. The validation and test sets remain the same.

\textbf{Commit Intent Classification} seeks to infer the purpose of a commit to a code change, for example, Bug, from its textual and code context. Accurate intent labels help reviewers prioritize urgent fixes, generate release notes, and support traceability throughout the CI/CD pipeline~\cite{math12071012}. 
For our experiments, we use the dataset in~\cite{WANG2021106408}, which contains commits whose titles, descriptions, changed-file paths, and diff hunks are manually annotated with one or more of seven intents: Bug, Refactor, Deprecation, Feature, Merge, Resource, and Test. 
To insert noise, we randomly replace the labels with a different intent. After filtering, we split the dataset into train, validate, and test using an 8:1:1 ratio, resulting in 700, 85, and 88 in the train, validate, and test datasets, respectively.
% This task provides a realistic yet controlled benchmark for testing how well \textsc{MANTRA}, applied to the five studied models can learn meaningful decision boundaries while dynamically adapting to corrupted supervision.

In both datasets, noisy data is replaced with the same ratio of clean data, thus, the dataset statistics remain the same.

\subsection{Models}
We selected five models of PTMs and LLMs selected in our study, informed by a previous work~\cite{comparison2023}, to assess MANTRA's effectiveness. \textbf{CodeBERT}~\cite{CodeBERT} is an encoder-only Transformer model that has demonstrated strong results across code-related tasks~\cite{Ahmed, codet5+} and is used in previous SE noise label learning studies~\cite{wang2024empirical}. \textbf{CodeT5+}~\cite{codet5+} designed for code understanding and generation. \textbf{CodeT5+}~\cite{codet5+} supports encoder-only, decoder-only, and encoder-decoder architectural, achieving state-of-the-art performance on more than 20 code benchmarks at the time of its release. \textbf{\codellama}~\cite{codellama} is a well-known series of large language models tailored specifically for code understanding and generation across multiple programming languages, built on the Transformer architecture~\cite{codellama}. \textbf{StarCoder2}~\cite{starcoder2} is a widely-used LLM for code, trained on The Stack v2, across multiple programming languages, and archived encouraging result. \textbf{Qwen2.5-Coder}~\cite{qwen2.5} is the latest code-specialized branch of the Qwen2.5 family and has performance gains in code intelligence and is studied in various SE research \cite{10.1145/3747588,10.1145/3641289}.
All five models are open source with checkpoints available from HuggingFace, and detailed specifications are provided in Table~\ref{tab:model_config}.

\begin{table}[t]
\centering
\caption{Pre-trained backbones and fine-tuning setup used in this study.  
         All checkpoints were obtained from HuggingFace and are open-source. }
\label{tab:model_config}
\begin{tabular}{@{}lccc@{}}
\toprule
\textbf{Model} & \textbf{\#Params} & \textbf{Context} & \textbf{FT method} \\
\midrule
\codebert   & 0.12 B  & 514     & Full FT \\
\ct         & 0.22 B  & 512     & Full FT \\
\codellama  & 7.0 B   & 16 384  & LoRA (r=16, $\alpha=16$) \\
\starcoder  & 7.0 B   & 16 384  & LoRA (r=16, $\alpha=16$) \\
\qwen       & 7.61 B  & 32 768  & LoRA (r=16, $\alpha=16$) \\
\bottomrule
\end{tabular}
\end{table}

\subsection{Evaluation Metrics} 
We explain the evaluation metrics used for each task in this section.
\subsubsection{Code Summarization}
We employ \textbf{BLEU-4} (Bilingual Evaluation Understudy for four-gram matches) as the primary metric to evaluate our code summarization approach. BLEU-4 quantifies how closely a generated summary aligns with a reference summary by comparing overlapping word sequences (\textit{n}-grams) of lengths 1 to 4. 

% Specifically:

% \begin{itemize}
%     \item \textbf{N-gram Overlap:} The metric calculates precision for each \textit{n}-gram length (1-gram, 2-gram, 3-gram, and 4-gram). A higher overlap between the generated text and the reference implies greater lexical similarity.

%     \item \textbf{Geometric Mean:} BLEU-4 combines these individual \textit{n}-gram precisions into a single score through a geometric mean, ensuring that the final metric balances short and long-phrase matches.

%     \item \textbf{Brevity Penalty:} To discourage overly short summaries that might artificially inflate precision, BLEU-4 applies a penalty if the generated summary is significantly shorter than the reference.
% \end{itemize}

% A higher BLEU-4 score thus indicates more accurate and fluent summaries, making it a well-suited metric for evaluating the lexical fidelity of automatically generated code summaries.

\subsubsection{Code Commit intention classification}
We employ the \textbf{micro averaged F\textsubscript{1} score} (Micro-F1) to evaluate our multilabel commit intent classification. Micro F1 score aggregates the total true positives (TP), false positives (FP), and false negatives (FN) over all classes and examples, then computes a single precision and recall before taking their harmonic mean:

For class \(i\) we define:

\(\mathrm{Precision}_{i}=TP_{i}/(TP_{i}+FP_{i})\) 
and \(\mathrm{Recall}_{i}=TP_{i}/(TP_{i}+FN_{i})\).  

Aggregating counts over all \(n\) classes gives the micro-averaged metrics:

\(\mathrm{Precision}_{\text{micro}}=\bigl(\sum_{i=1}^{n}TP_{i}\bigr)\big/\bigl(\sum_{i=1}^{n}(TP_{i}+FP_{i})\bigr)\) and  
\(\mathrm{Recall}_{\text{micro}}=\bigl(\sum_{i=1}^{n}TP_{i}\bigr)\big/\bigl(\sum_{i=1}^{n}(TP_{i}+FN_{i})\bigr)\).  
Their harmonic mean yields the micro-\(F_{1}\):

\[
F1_{\text{micro}}
=2\cdot\frac{\text{Precision}_{\text{micro}}\!\cdot\!\text{Recall}_{\text{micro}}}
            {\text{Precision}_{\text{micro}}+\text{Recall}_{\text{micro}}}
.
\]

Because it weights every individual prediction equally, Micro-F1 provides a single, unified measure of overall classification quality, particularly well-suited to imbalanced or multi-label settings, such as commit intent analysis.

\subsection{Experiment setup}
All experiments were run on a GPU node with 1 CPU core and an A100 40 GB GPU. We used the hyper-parameter configurations specified in the original model papers, training the models with a learning rate of 0.00005. 

In MANTRA, we choose to drop the samples after 3 epochs for code summarization and 5 epochs for commit intent classification. These numbers are set based on our observations from RQ1. We use the scikit-learn\footnote{\url{https://scikit-learn.org/stable/index.html}{https://scikit-learn.org/stable/index.html}} library for all GMM-related parts. 

%% file: sections/result.tex
\section{Results}
In this section, we will present the findings of our research questions, followed by an analysis of the observations. We present the results of each of the two tasks separately in the following. 

\subsection{Loss Trajectory and Convergence Behavior in Presence of Noise}

\subsubsection{Code Summarization}

Table~\ref{tab:combined_noise_levels} shows the average loss for code summarization for CodeBERT and CodeT5+, with various noise ratios over 10 epochs. 
% \FHF{Add a sentence that loss should decrease with more epochs, the sooner it happens, the sooner model converges. }
The results reveal consistent patterns in the loss trajectories of these models during training, with variations based on the level of noise introduced into the dataset. In all noise levels, the loss decreases over epochs. However, the rate of this reduction and the final loss values are strongly influenced by the level of noise, showing the detrimental effects of noise on training.

\begin{table}[h]
    \centering
    \resizebox{\columnwidth}{!}{
        \begin{tabular}{|c|c|c|c|c|c|c|c|c|c|c|c|}
            \hline
            \textbf{Noise} & \textbf{Type} & \textbf{1} & \textbf{2} & \textbf{3} & \textbf{4} & \textbf{5} & \textbf{6} & \textbf{7} & \textbf{8} & \textbf{9} & \textbf{10} \\
            \hline
            
            \multicolumn{12}{|c|}{\textbf{CodeBERT}} \\
            \hline
            \multirow{2}{*}{5\%} & Noise & 5.36 & 4.61 & 4.45 & 4.31 & 4.27 & 4.19 & 4.14 & 4.10 & 4.02 & 4.00 \\
                                 & clean & 4.86 & 3.42 & 3.01 & 2.80 & 2.59 & 2.44 & 2.32 & 2.20 & 2.14 & 2.05 \\
            \hline
            \multirow{2}{*}{10\%} & Noise & 5.39 & 4.59 & 4.33 & 4.24 & 4.17 & 4.10 & 4.02 & 4.00 & 3.93 & 3.90 \\
                                  & clean & 4.85 & 3.62 & 3.22 & 3.09 & 2.86 & 2.69 & 2.53 & 2.45 & 2.34 & 2.29 \\
            \hline
            \multirow{2}{*}{15\%} & Noise & 5.30 & 4.56 & 4.29 & 4.21 & 4.13 & 4.04 & 3.94 & 3.87 & 3.88 & 3.84 \\
                                  & clean & 4.97 & 3.74 & 3.42 & 3.20 & 3.00 & 2.85 & 2.62 & 2.55 & 2.55 & 2.40 \\
            \hline
            % \multirow{2}{*}{80\%} & Noise & 5.28 & 4.37 & 4.15 & 4.00 & 3.88 & 3.76 & 3.66 & 3.58 & 3.54 & 3.50 \\
            %                       & clean & 5.28 & 4.37 & 4.17 & 4.00 & 3.90 & 3.79 & 3.63 & 3.59 & 3.55 & 3.48 \\
            % \hline
            % \multirow{2}{*}{90\%} & Noise & 5.25 & 4.37 & 4.15 & 4.01 & 3.89 & 3.77 & 3.66 & 3.58 & 3.54 & 3.50 \\
            %                       & clean & 5.23 & 4.34 & 4.12 & 3.97 & 3.89 & 3.77 & 3.66 & 3.58 & 3.53 & 3.50 \\
            % \hline

            \multicolumn{12}{|c|}{\textbf{\ct}} \\
            \hline

            \multirow{2}{*}{5\%}
            & Noise
            & 1.74 & 0.57 & 0.55 & 0.54 & 0.55 & 0.52 & 0.55 & 0.51 & 0.55 & 0.54 \\
            & clean
            & 2.14 & 0.30 & 0.27 & 0.25 & 0.23 & 0.21 & 0.19 & 0.18 & 0.15 & 0.14 \\
            \hline
            
            % 10%
            \multirow{2}{*}{10\%}
            & Noise
            & 1.96 & 0.54 & 0.54 & 0.54 & 0.54 & 0.52 & 0.54 & 0.53 & 0.54 & 0.53 \\
            & clean
            & 2.05 & 0.34 & 0.31 & 0.31 & 0.31 & 0.29 & 0.27 & 0.28 & 0.24 & 0.23 \\
            \hline
            
            % 15%
            \multirow{2}{*}{15\%}
            & Noise
            & 1.82 & 0.55 & 0.54 & 0.53 & 0.54 & 0.52 & 0.53 & 0.54 & 0.53 & 0.54 \\
            & clean
            & 2.08 & 0.36 & 0.35 & 0.33 & 0.32 & 0.31 & 0.33 & 0.33 & 0.30 & 0.29 \\
            \hline
            
            % % 80%
            % \multirow{2}{*}{80\%}
            % & Noise
            % & 1.67 & 0.54 & 0.53 & 0.53 & 0.53 & 0.54 & 0.54 & 0.53 & 0.53 & 0.53 \\
            % & clean
            % & 1.71 & 0.52 & 0.51 & 0.51 & 0.51 & 0.52 & 0.52 & 0.52 & 0.52 & 0.51 \\
            % \hline
            
            % % 90%
            % \multirow{2}{*}{90\%}
            % & Noise
            % & 1.63 & 0.54 & 0.54 & 0.54 & 0.53 & 0.54 & 0.53 & 0.54 & 0.53 & 0.53 \\
            % & clean
            % & 1.66 & 0.53 & 0.52 & 0.52 & 0.52 & 0.52 & 0.52 & 0.52 & 0.52 & 0.52 \\
            % \hline
            
        \end{tabular}
    }
    \caption{Average loss for code summarization for CodeBERT and \ct~at different noise levels (5\%, 10\%, 15\%) randomly inserted into the dataset. Each row compares the loss for noisy data (Noise) and non-noisy data (clean) across epochs, shown with numbers 1--10. clean means 0\% noise.}
    \label{tab:combined_noise_levels}
\end{table}

The loss reduction for noisy data is slower compared to clean data. For instance, with 5\% noise, the loss at epoch 1 for noisy data is 5.36 for \codebert, while for clean data it is 4.86. This gap widens by epoch 10, where the losses are 4.00 and 2.05, respectively. Same trend is observed for 10\% and 15\% noise levels, and for \ct~model (though in the first epoch, \ct~has higher loss for clean data). This widening gap over epochs indicates that while the model can still converge on noisy data, the process becomes more challenging due to the presence of noise. The model requires additional epochs to effectively minimize loss, suggesting that even small amounts of noise can impede learning.
Overall, the clean data has lower loss in the last epoch compared to the noisy data. 
% Considering the loss of noisy vs. clean data, we observe that the model converges faster (i.e., has lower loss in the last epoch).
% , specially after epoch 1, where the loss in the second epoch decreases significantly compared to the first epoch.

% \FHF{Maybe remove this paragraph:}
% \textbf{Convergence Behavior and Epoch Dynamics}
% In early epochs, the impact of noise appears less pronounced, particularly at lower noise levels. For example, at 5\% noise, the loss gap between noisy and non-noisy data in epoch 1 is around 0.50 for CodeBERT. However, this gap grows significantly as training progresses, reaching 1.95 by epoch 10. This suggests that the model processes noisy and non-noisy data early in training more similarly, but the cumulative effects of noise manifest as the optimization deepens. 
% \FHF{This is not correct:}
% In contrast, at higher noise levels, the loss gap between noisy and non-noisy datasets is negligible from the start, with both showing minimal improvement after epoch 6. This stagnation in learning indicates a plateau in convergence caused by overwhelming noise.
% \FHF{Also, it does not hold for the codet5+ model. }

\textbf{Finding 1:} \textit{For code summarization, the presence of noise leads to higher final losses and slower convergence for PTMS. }

% The presence of noise significantly affects the loss trajectory and convergence behavior of code-pre-trained models during training for code summarization. The loss trajectories for noisy data are characterized by slower convergence and higher final losses.

% , particularly at low to moderate noise levels.

% Line Plot
% \begin{figure}[H]
%     \centering
%     \includegraphics[width=\columnwidth]{plots/r_line.pdf}
%     \caption{Loss Across Epochs (Subplots) under random insertion for \codebert}
%     \label{fig:r_line}
% \end{figure}
% \begin{figure}[H]
%     \centering
%     \includegraphics[width=\columnwidth]{plots/p_line.pdf}
%     \caption{Loss Across Epochs (Subplots) under random insertion for codeT5+}
%     \label{fig:p_line}
% \end{figure}

\begin{figure}[!]
  \centering
  \begin{subfigure}[t]{0.49\columnwidth}
    \centering
    \includegraphics[width=\linewidth]{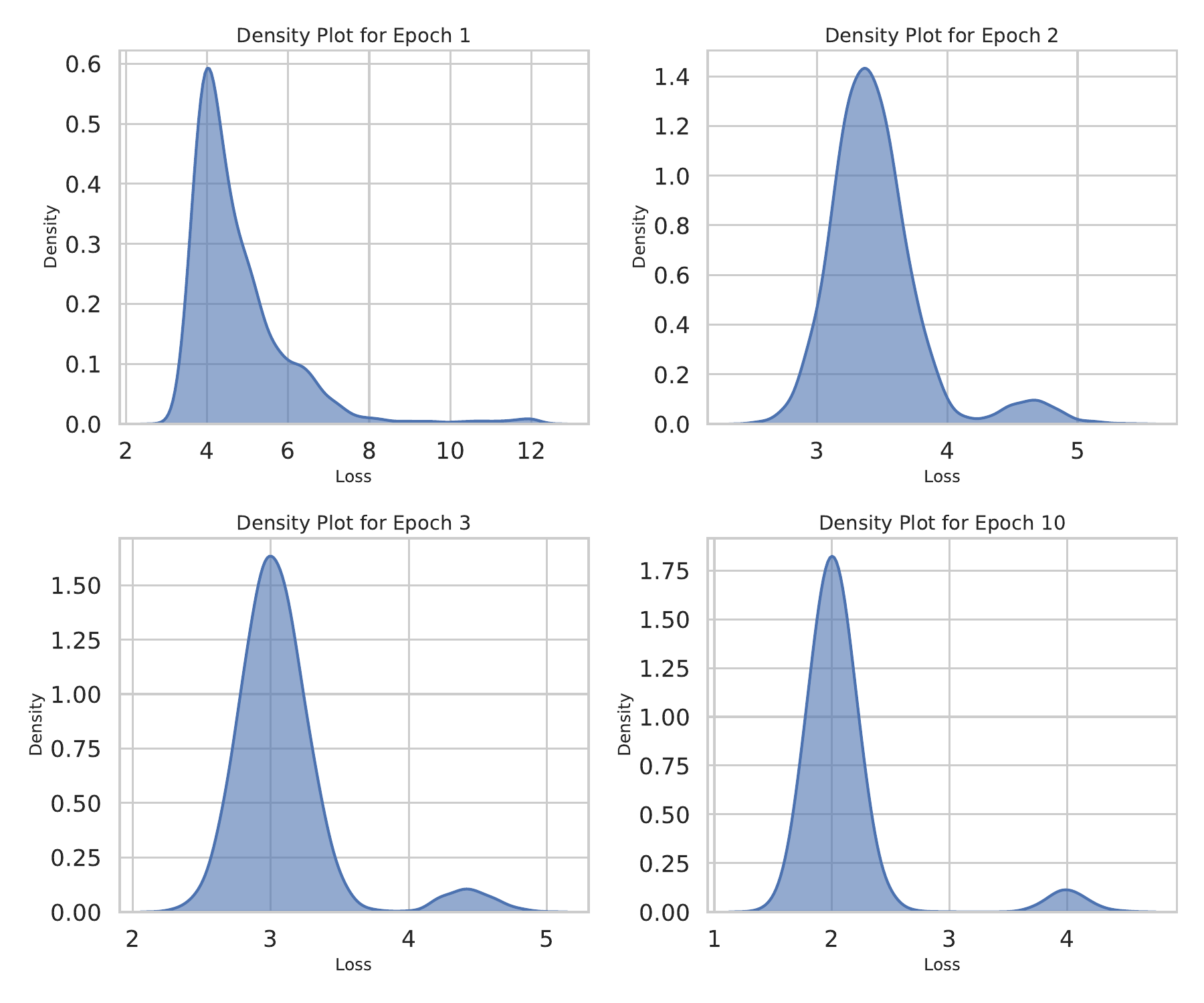}
    \caption{Distribution of loss for \codebert\ (5\%, epochs 1, 2, 3, 10)}
    \label{fig:end_box-codebert}
  \end{subfigure}\hfill
  \begin{subfigure}[t]{0.49\columnwidth}
    \centering
    \includegraphics[width=\linewidth]{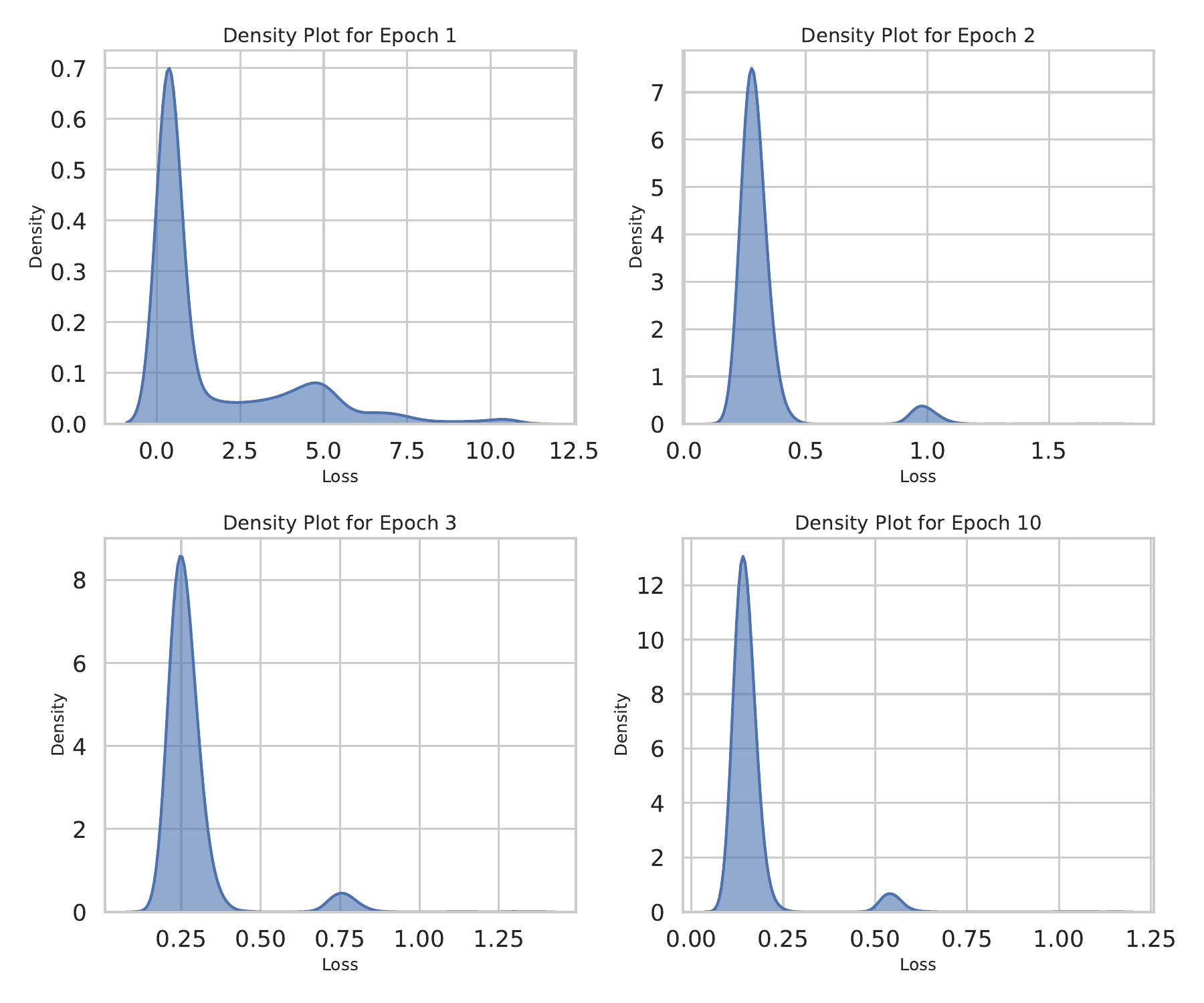}
    \caption{Distribution of loss for \ct\ (5\%, epochs 1, 2, 3, 10)}
    \label{fig:end_box-codet5}
  \end{subfigure}
  \caption{Distribution of loss for \codebert\ and \ct\ at 5\% across epochs 1, 2, 3, and 10.}
  \label{fig:end_box-both}
\end{figure}

\textbf{Loss Density Plots for Code Summarization.}
The density plots in Figures~\ref{fig:end_box-codebert} and ~\ref{fig:end_box-codet5} illustrate the evolution of the loss distribution across different epochs for \codebert and \ct, highlighting the models' optimization process, over different epochs in presence of 5\% noise. In the first epoch, the distribution is wide and relatively flat, indicating high variance in loss values. 
% as the model begins training with \FHF{unoptimized predictions??}. 
By epoch 2, the distribution becomes more peaked, reflecting reduced variance and a shift toward lower loss values as the model starts learning meaningful patterns. This trend continues in later epochs, where the density curve narrows further, and the peak moves closer to lower loss values.
% , signaling improved convergence. 
By epoch 10, the loss distribution is sharply peaked, with most values concentrated near the lower end of the scale, indicating that the model has effectively minimized loss for the majority of samples. This progressive narrowing and shifting of the density plots demonstrates the model's ability to reduce variance and improve consistency across samples during training. Any remaining width in the density plot or secondary peaks at later epochs could suggest the influence of noisy samples or outliers, which the model may struggle to optimize fully.

\textbf{Finding 2:} \textit{For both PTMs, we observe two distinct distributions after the second epoch in their loss density plots for code summarization, which we relate to losses for non-noisy and noisy data. }

% \begin{figure}[H]
%     \centering
%     \includegraphics[width=\columnwidth]{plots/combined_density_plots_t5p.pdf}
%     \caption{Distribution of loss for \ct for 5\% error for batch 1, 2, 3, and 10}
%     \label{fig:p_end_box}
% \end{figure}

% \begin{figure}[H]
%     \centering
%     \includegraphics[width=\columnwidth]{plots/combined_density_plots_t5p.pdf}
%     \caption{Distribution of loss for \codellama for 5\% error for batch 1, 2, 3, and 10}
%     \label{fig:p_end_box}
% \end{figure}

% \begin{figure}[H]
%     \centering
%     \includegraphics[width=\columnwidth]{plots/combined_density_plots_t5p.pdf}
%     \caption{Distribution of loss for \starcoder for 5\% error for batch 1, 2, 3, and 10}
%     \label{fig:p_end_box}
% \end{figure}

% \begin{figure}[H]
%     \centering
%     \includegraphics[width=\columnwidth]{plots/combined_density_plots_t5p.pdf}
%     \caption{Distribution of loss for \qwen for 5\% error for batch 1, 2, 3, and 10}
%     \label{fig:p_end_box}
% \end{figure}

\begin{figure*}[h]
  \centering
  \includegraphics[width=\textwidth]{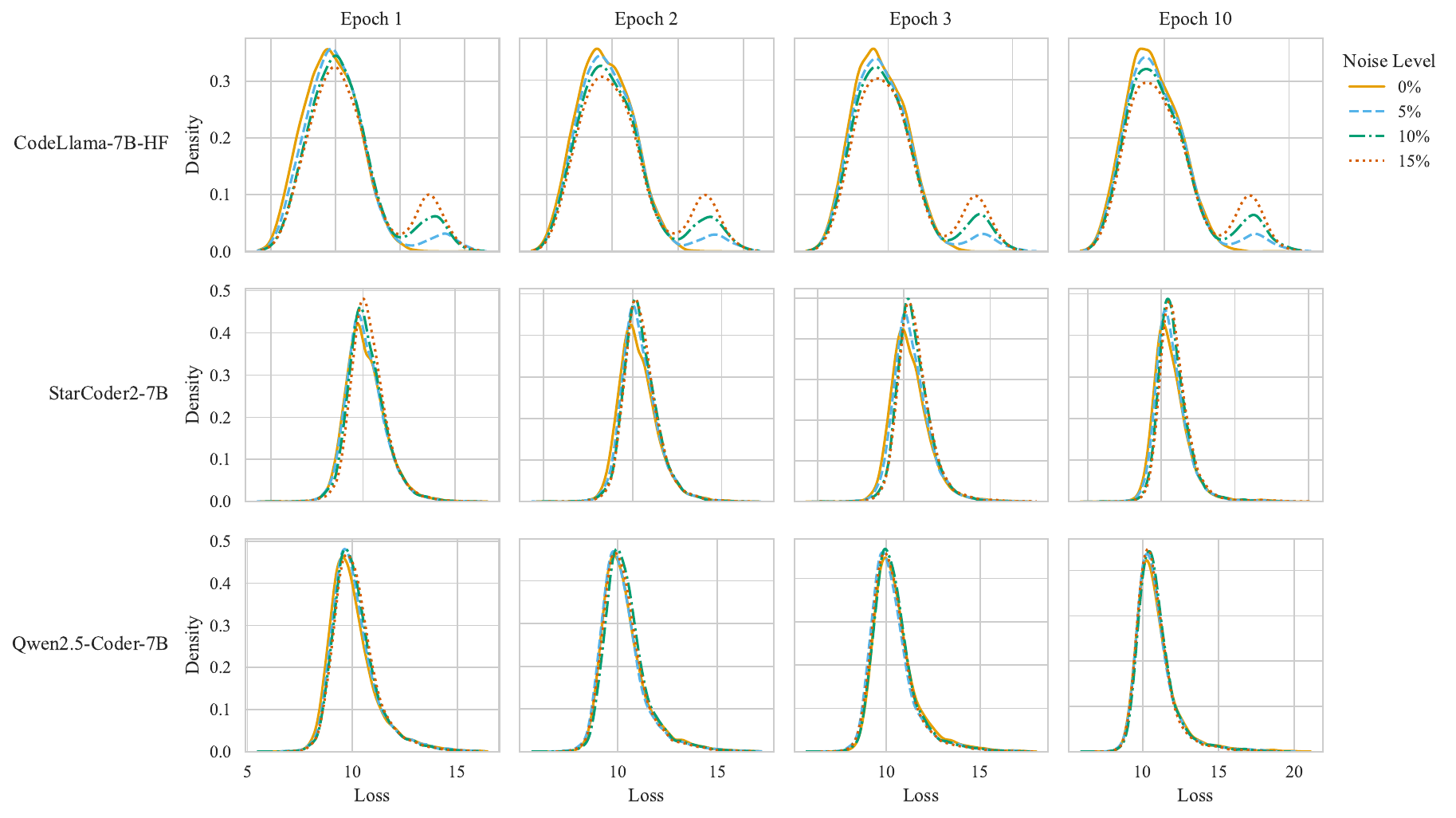}
  \caption{Loss density distributions by model, epoch, and noise level for code summarization. Each subplot has independent axes.}
  \label{fig:loss-density-cs}
  \vspace{-0.5em}
\end{figure*}

The plots in Figure~\ref{fig:loss-density-cs} provide insights into the behaviors of the evaluated LLMs for code summarization under varying noise levels and training epochs. 
The clean data is shown with \textcolor{orange}{\rule[0.5ex]{1em}{0.5pt}}, while data with 5\%, 10\%, and 15\% inserted noise are shown in 
\tikz[baseline] \draw[dashed, thick, blue] (0ex,0.5ex) -- +(3ex,0);,
\tikz[baseline] \draw[dashed, thick, darkgreen] (0ex,0.5ex) -- +(3ex,0);, and 
\tikz[baseline] \draw[dashed, thick, red] (0ex,0.5ex) -- +(3ex,0);, respectively. 
Firstly, \codellama~clearly exhibits a dual distribution in its loss values under higher noise conditions (10\% and 15\%), indicating a distinct separation between clean and noisy samples during training. This separation suggests that \codellama~effectively learns to distinguish genuine data from corrupted samples, thereby forming a clear boundary in the loss space. 
% The effectiveness of such a phenomenon is reinforced by the performance improvements presented in the final results table, which demonstrates notable gains when applying MANTRA.

In contrast, \starcoder~and \qwen~show less differentiation in their loss distributions. Both models exhibit a broader tail in higher noise scenarios and peak shifting to the left for the clean data, yet fail to demonstrate clear bi-distributions. This implies a relatively limited capability in discriminating noisy samples based solely on loss values. Interestingly, the performance drop of \starcoder~with noise inserted remains marginal for code summarization (See Table~\ref{tab:merged_noise_performance}), suggesting the presence of robustness within this model to noise interference for this task. Conversely, \qwen~exhibits significant performance degradation under noise conditionss (See Table~\ref{tab:merged_noise_performance}).
% , but sees substantial recovery when integrating MANTRA, reinforcing the importance of adaptive noise-handling strategies in preserving model robustness. 
A plausible explanation for this behaviour is that \starcoder~may inherently has a more robust internal representation, or better learning ability on noisy data, due to its pre-training process, which may include diverse and highly noisy data. This robustness could allow the model to generalize better across different data qualities without the need to explicitly distinguish noisy samples in the loss space.

\textbf{Finding 3:} \textit{Overall, these observations highlight the varying sensitivities of LLMs to noise for code summarization, underscoring the necessity of tailored approaches in noise mitigation.
}
% \FHF{unclear: what do you mean?}
% \codellama's clear bimodal behavior aligns with theoretical expectations that effective model training should distinctly separate clean and noisy data points over successive epochs. In contrast, \starcoder and \qwen subtle differentiation may indicate either inherent robustness or limited sensitivity to noise. 

% Overall, these results affirm the efficacy and necessity of dynamically identifying and removing noisy samples during training, which contributes significantly to enhancing generalization performance, especially in highly noisy training scenarios.

\subsubsection{Code Commit Intent Classification}

\begin{figure*}[h]
  \centering
  \includegraphics[width=\textwidth]{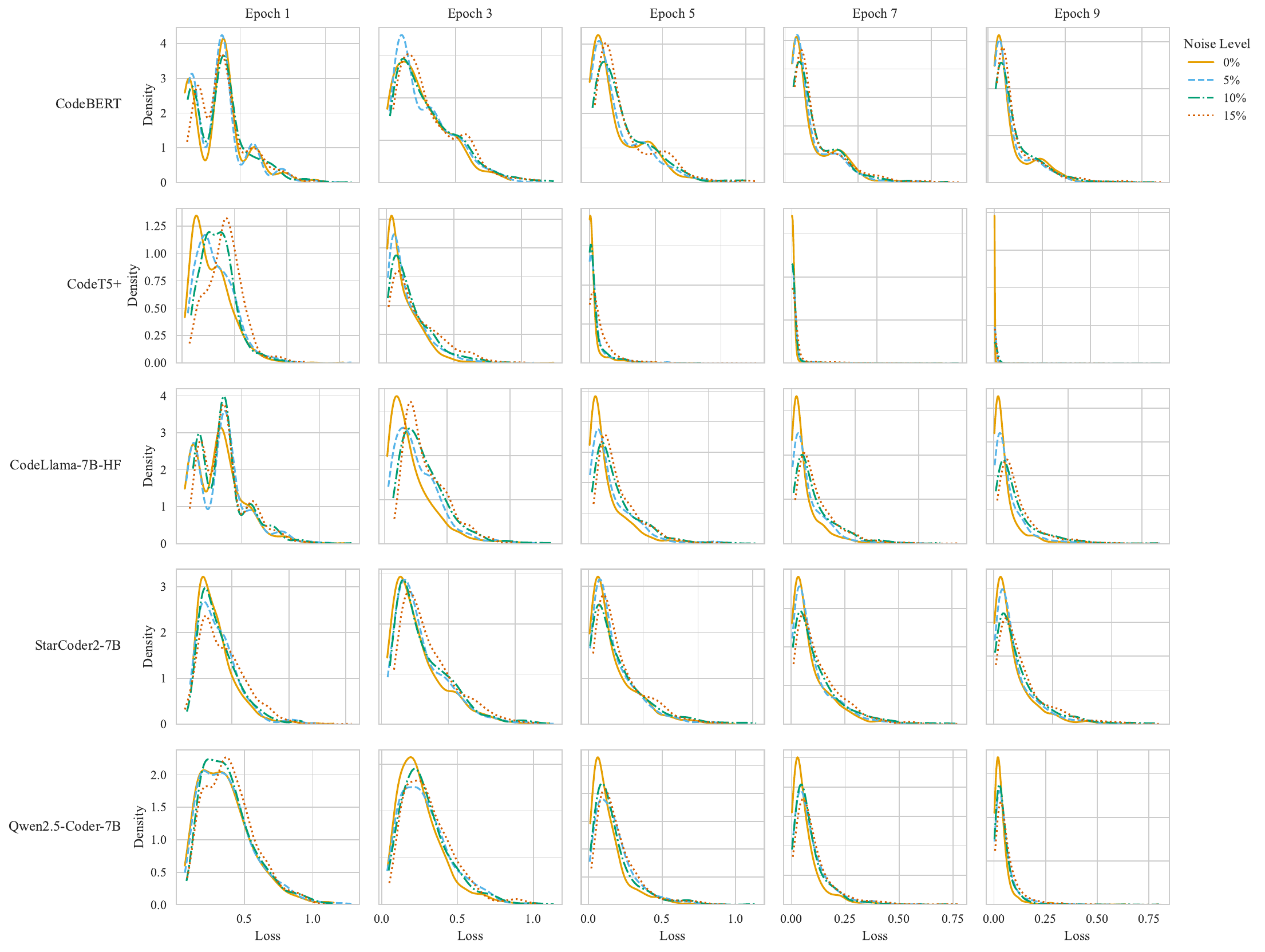}
  \caption{Loss density distributions by model, epoch, and noise level for code commit intent classification. Each subplot has independent axes.}
  \label{fig:loss-density-commit}
  \vspace{-0.5em}
\end{figure*}

\begin{figure}[h]
    \centering
    \includegraphics[width=0.9\columnwidth]{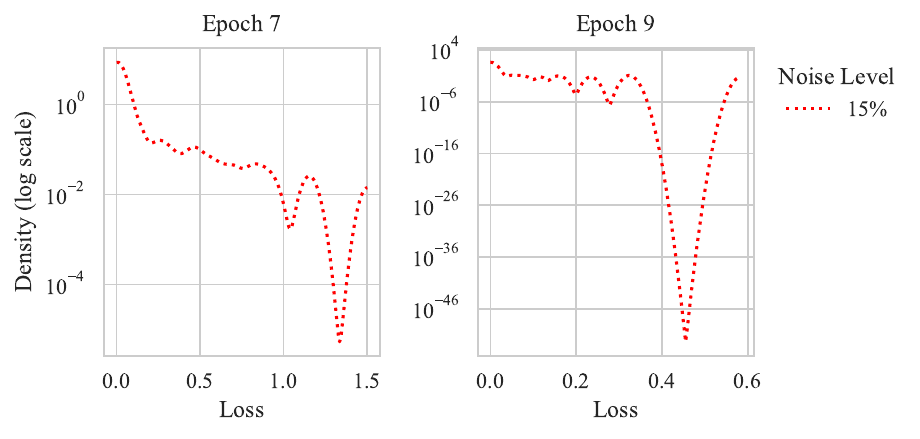}
    \caption{Loss density distributions for \ct~at epochs 7 and 9 under 15\% noise insertion on a log scale.}
    \label{fig:codet5p_loss}
\end{figure}

Figure~\ref{fig:loss-density-commit} presents the loss density distributions of the five studied models for commit intent classification, over epochs 1, 3, 5, 7, and 9, with different noise levels: 0 (clean data), 5\%, 10\%, and 15\%. The clean data is shown with \textcolor{orange}{\rule[0.5ex]{1em}{0.5pt}}, while data with 5\%, 10\%, and 15\% noise are shown in 
\tikz[baseline] \draw[dashed, thick, blue] (0ex,0.5ex) -- +(3ex,0);,
\tikz[baseline] \draw[dashed, thick, darkgreen] (0ex,0.5ex) -- +(3ex,0);, and 
\tikz[baseline] \draw[dashed, thick, red] (0ex,0.5ex) -- +(3ex,0);, respectively. 
We can see that the higher the noise level, the more difficult it is for the model to fit and converge in each training phase, especially in the last epoch. 
In epoch 9, the loss distribution of 0\% noise samples has been clustered into a very sharp and high peak, meaning that all clean samples are firmly fitted to the same low-loss interval by the model, which is most obvious in \ct, where the height of the peak is so high that other noise levels are almost invisible. 
As the percentage of noise increases, the corresponding density's peak decreases and becomes progressively wider to the right, with the right tail becoming more pronounced. 
In other words, it is easier for clean data to find a low-loss solution in the parameter space, while the noise samples can neither share the same loss interval nor be highly aggregated in the distribution due to the conflict with the true distribution.
\codellama~in Figure~\ref{fig:loss-density-commit}, shows clear noise effects at a very early stage (epoch 3). \starcoder~shows highly consistent convergence after the fifth round of training, regardless of the amount of noise.
Even with 15\% noise samples, it still demonstrates a wider tail.
\qwen~also shows persistent tail thickness, reflecting their strong, but more gradual ability to learn in the presence of noise.
% , a phenomenon also evident in their test-set performance, which we will discuss in result of RQ2.
These observations mostly align with loss density plots for code summarization (Figures~\ref{fig:end_box-codebert}, \ref{fig:end_box-codet5}, and \ref{fig:loss-density-cs}), where \codebert, \ct and \codellama~show more clear distinction among the noisy and clean data, while \starcoder and \qwen~demonstrate less difference. 

\textbf{Finding 4:} \textit{Overall, for commit intent classification, all models experience progressively wider tails in later epochs as noise continues to increase, suggesting that the models have difficulty converging the noisy samples alongside the clean ones. }

\subsection{Performance of MANTRA}

% \begin{table}[h]
% \centering
% \begin{tabular}{|c|c|c|c|c|}
% \hline
%  & \textbf{Reported} & \textbf{5\%}& \textbf{10\%}& \textbf{15\%} \\
% \hline
% \codebert & 18.24 & 18.19 & 18.01 & 17.69\\
% \codebert dropout & 18.3 & 18.21 & 18.18 & 18.14\\
% \hline
% \ct & 17.71  & 17.25 & 17.05 & 16.86 \\
% \ct dropout &  17.51  & 18.39 & 18.13 & --\\
% \hline
% \codellama &  14.82  & 12.51 & 11.48 & 11.45\\
% \codellama dropout &  14.82  & -- & -- & --\\
% \hline
% \starcoder &  17.49  & 17.2 & 17.02 & 16.67\\
% \starcoder dropout &  17.49  & -- & -- & --\\
% \hline
% \qwen &  17.49  & 17.2 & 17.02 & 16.67\\
% \qwen dropout &  17.49  & -- & -- & --\\
% \hline
% \end{tabular}
% \caption{Performance for dropout and insert noise}
% \end{table}
\subsubsection{Code Summarization}

Table~\ref{tab:merged_noise_performance} reports the BLEU-4 score for all models studied in code summarization. 
\codebert performs best with a BLEU-4 of 18.24 on the clean data, but drops to 17.69 at 15\% noise, a drop of 0.55. With MANTRA, the score rises to 18.30 at no noise, and more importantly, drops only to 18.14 at 15\% noise, narrowing the drop to 0.16. \ct~dropped from 17.71 to 16.86 (down 0.85) under fully fine-tuning, instead slightly increased the score to 18.39 at 5\% noise and kept it at 17.77 at 15\% noise with MANTRA, a drop of only 0.24 from the clean column. For \codellama, the BLEU-4 at 15\% noise dropped directly from 14.82 to 11.45, which is an alarming drop; however, with MANTRA, only drops from 12.35 to 11.24, which is a drop of 1.11. The drop of \starcoder~%fine-tuning with LoRA 
is mild (14.66 to 13.60, a drop of 1.06), but after adding MANTRA, it not only rises back to 14.52 and 14.60 at 5\% and 10\% noise, respectively, but also increases to 14.52 and 14.61 at 5\% and 10\% noise, and a similar performance of 14.49 with 15\% noise using MANTRA. \qwen~is the most sensitive model to noise, where the score plummets to 2.66 at 15\% noise% with LoRA fine-tuning
, but with MANTRA, it recovers to 9.51 at worst, which is a narrower drop and significantly improves the robustness.

The general observation comparing the models' performance with and without noise is similar to RQ1; the higher the proportion of noise, the greater the decline, which is different for various models. We observe a more significant decline in \qwen.

\subsubsection{Code Commit Intent Classification}

% \begin{table}[ht]
% \centering
% \caption{F1 scores of different models on commit intent classification under varying label noise levels (0\%, 5\%, 10\%, 15\%), with and without dropout \JW{should we merge Table 3 and Table 4 to save space?}}
% \label{tab:commit_intent_classification}
% \begin{tabular}{lcccc}
% \toprule
% \textbf{Model (Method)} & \textbf{0\%} & \textbf{5\%} & \textbf{10\%} & \textbf{15\%} \\
% \midrule
% \multicolumn{5}{l}{\itshape \codebert} \\
% \quad fully Fine-tuning & 69.4\% & 64.7\% & 49.5\% & 56.8\% \\
% \quad fully Fine-tuning +  MANTRA    & 67.9\% & 66.1\% & 56.9\% & 56.9\% \\
% \midrule
% \multicolumn{5}{l}{\itshape \ct} \\
% \quad fully Fine-tuning & 68.2\% & 66.4\% & 60.6\% & 61.3\% \\
% \quad fully Fine-tuning +  MANTRA    & 67.9\% & 69.3\% & 67.6\% & 67.5\% \\
% \midrule
% \multicolumn{5}{l}{\itshape \codellama} \\
% \quad LoRA & 68.8\% & 65.0\% & 60.0\% & 59.6\% \\
% \quad LoRA +  MANTRA   & 72.0\% & 76.4\% & 72.0\% & 69.2\% \\
% \midrule
% \multicolumn{5}{l}{\itshape \starcoder} \\
% \quad LoRA & 68.5\% & 70.3\% & 62.4\% & 58.6\% \\
% \quad LoRA +  MANTRA    & 64.2\% & 63.2\% & 65.1\% & 64.8\% \\
% \midrule
% \multicolumn{5}{l}{\itshape \qwen} \\
% \quad LoRA & 58.7\% & 51.6\% & 51.3\% & 44.8\% \\
% \quad LoRA +  MANTRA    & 61.1\% & 55.1\% & 55.3\% & 52.3\% \\
% \bottomrule
% \end{tabular}

% \end{table}

\begin{table}[ht]
  \centering
  \setlength\tabcolsep{3pt}
  \small
  \caption{BLEU scores for code summarization (CS) and F1 scores for commit intent classification (F1) under varying noise levels (0\%, 5\%, 10\%, 15\%), comparing baseline and MANTRA-enhanced methods.}
  \label{tab:merged_noise_performance}
  \small
  \begin{tabular}{l*{4}{cc}}
    \toprule
    \multirow{2}{*}{\textbf{Model (Method)}} &
      \multicolumn{2}{c}{\textbf{0\%}} &
      \multicolumn{2}{c}{\textbf{5\%}} &
      \multicolumn{2}{c}{\textbf{10\%}} &
      \multicolumn{2}{c}{\textbf{15\%}} \\
    \cmidrule(lr){2-3}\cmidrule(lr){4-5}\cmidrule(lr){6-7}\cmidrule(lr){8-9}
    & \textbf{CS} & \textbf{F1} & \textbf{CS} & \textbf{F1} & \textbf{CS} & \textbf{F1} & \textbf{CS} & \textbf{F1} \\
    \midrule
    \multicolumn{9}{l}{\itshape CodeBERT} \\
    \quad full FT           & 18.24 & 69.4\% & 18.19 & 64.7\% & 18.01 & 49.5\% & 17.69 & 56.8\% \\
    \quad MANTRA & 18.30 & 67.9\% & 18.21 & 66.1\% & 18.18 & 56.9\% & 18.14 & 56.9\% \\
    \midrule
    \multicolumn{9}{l}{\itshape CodeT5+} \\
    \quad full FT           & 17.71 & 68.2\% & 17.25 & 66.4\% & 17.05 & 60.6\% & 16.86 & 61.3\% \\
    \quad MANTRA & 17.51 & 67.9\% & 18.39 & 69.3\% & 18.13 & 67.6\% & 17.77 & 67.5\% \\
    \midrule
    \multicolumn{9}{l}{\itshape CodeLlama-7B} \\
    \quad LoRA                        & 14.82 & 68.8\% & 12.51 & 65.0\% & 11.48 & 60.0\% & 11.45 & 59.6\% \\
    \quad MANTRA              & 12.35 & 72.0\% & 11.77 & 76.4\% & 11.36 & 72.0\% & 11.24 & 69.2\% \\
    \midrule
    \multicolumn{9}{l}{\itshape StarCoder2-7B} \\
    \quad LoRA                        & 14.66 & 68.5\% & 14.03 & 70.3\% & 13.84 & 62.4\% & 13.60 & 58.6\% \\
    \quad MANTRA              & 14.46 & 64.2\% & 14.52 & 63.2\% & 14.61 & 65.1\% & 14.49 & 64.8\% \\
    \midrule
    \multicolumn{9}{l}{\itshape Qwen2.5-Coder-7B} \\
    \quad LoRA                        & 10.94 & 58.7\% &  6.66 & 51.6\% &  3.02 & 51.3\% &  2.66 & 44.8\% \\
    \quad MANTRA              & 12.01 & 61.1\% & 10.76 & 55.1\% & 10.45 & 55.3\% &  9.51 & 52.3\% \\
    \bottomrule
  \end{tabular}
\end{table}

Table~\ref{tab:merged_noise_performance} presents the F1 scores for all models with and without MANTRA. The F1 scores of all models drop significantly as the level of noise increases from to 15\%.
%, and the decrease is more pronounced as the proportion of noise increases
Most models' scores drop by 8-15 percentage points at 15\% noise compared to 0\% noise. We see the same trend in our results in the loss density distribution (RQ1). The noisy samples broaden the distribution and lower the peaks, making it difficult for the models to converge to the same low-loss interval as the clean samples.

When comparing the models, \codebert~achieves the best F1 of 69.4 without noise, but falls to 56.8 with 15\% noise, a drop of 12.6 points. Adding MANTRA narrows the drop to 11.0 points. \ct's performance is much smoother: without MANTRA, the drop from 0\% to 15\% is 7.6 points (from 68.2 to 60.6), while with MANTRA, the gap almost disappears, with a drop of only 0.4 points (from 67.9 to 67.5), which suggests that in the moderately noisy scenarios, MANTRA significantly improves the robustness of the model to noise. On the other hand, \codellama~ shows excellent robustness: without MANTRA, it drops 9.2 points (from 68.8 to 59.6), while with MANTRA, it drops only 2.8 points (from 72.0 to 69.2). Even with 15\% noise level, the performance is better than the 0\% data, which improves the overall performance and reduces the performance fluctuation caused by noise.

For \starcoder, the F1 drops from 68.5 to 58.6 without MANTRA at 15\%, but with MANTRA, the performance is improved a bit, being 64.8 under 15\% noise.~\qwen, is the most sensitive to noise. Without MANTRA, it drops by 13.9 points (from 58.7 to 44.8). With MANTRA, the drop is reduced to 9.4 points (from 61.1 to 52.3), showing that even though the model's ability to converge to noise is relatively weak, MANTRA still improves its performance and stability under noisy environments.

In summary, the results show high noise hinders convergence, but MANTRA’s dynamic removal of noisy samples yields smoother, more consistent performance across all noise levels. \ct~and \codellama~maintain almost zero degradation after the dropout, \qwen~is still stable in strong noise, and even in \codebert~ ~, we observe that the performance drop caused by the noise reduces significantly due to the introduction of MANTRA. These results not only validate the effect of noise on convergence that we see in the loss density distribution but also show that, in practice, in order to cope with noisy datasets, MANTRA is a reasonable choice that can significantly improve the stability of the training of the models. 
% and the ultimate generalization ability.}

%% file: sections/discussion.tex
\section{Discussion}

\begin{figure}[H] 
\centering 
\includegraphics[width=0.9\textwidth] 
{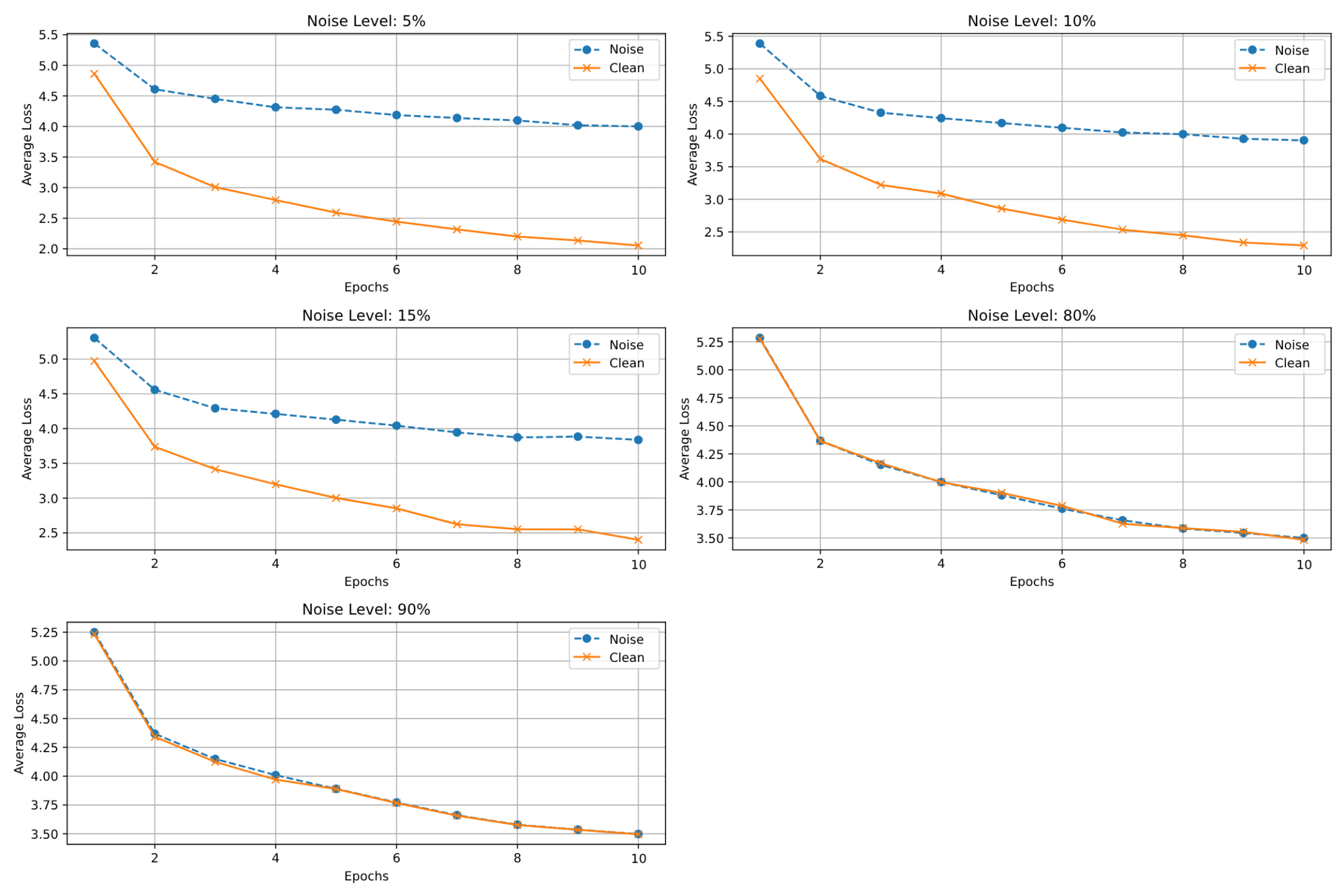} 
\caption{Loss Across Epochs (Subplots) for \codebert} \label{fig:r_line} 
\end{figure} 

\begin{figure}[b] 
\centering 
\includegraphics[width=0.9\columnwidth]
{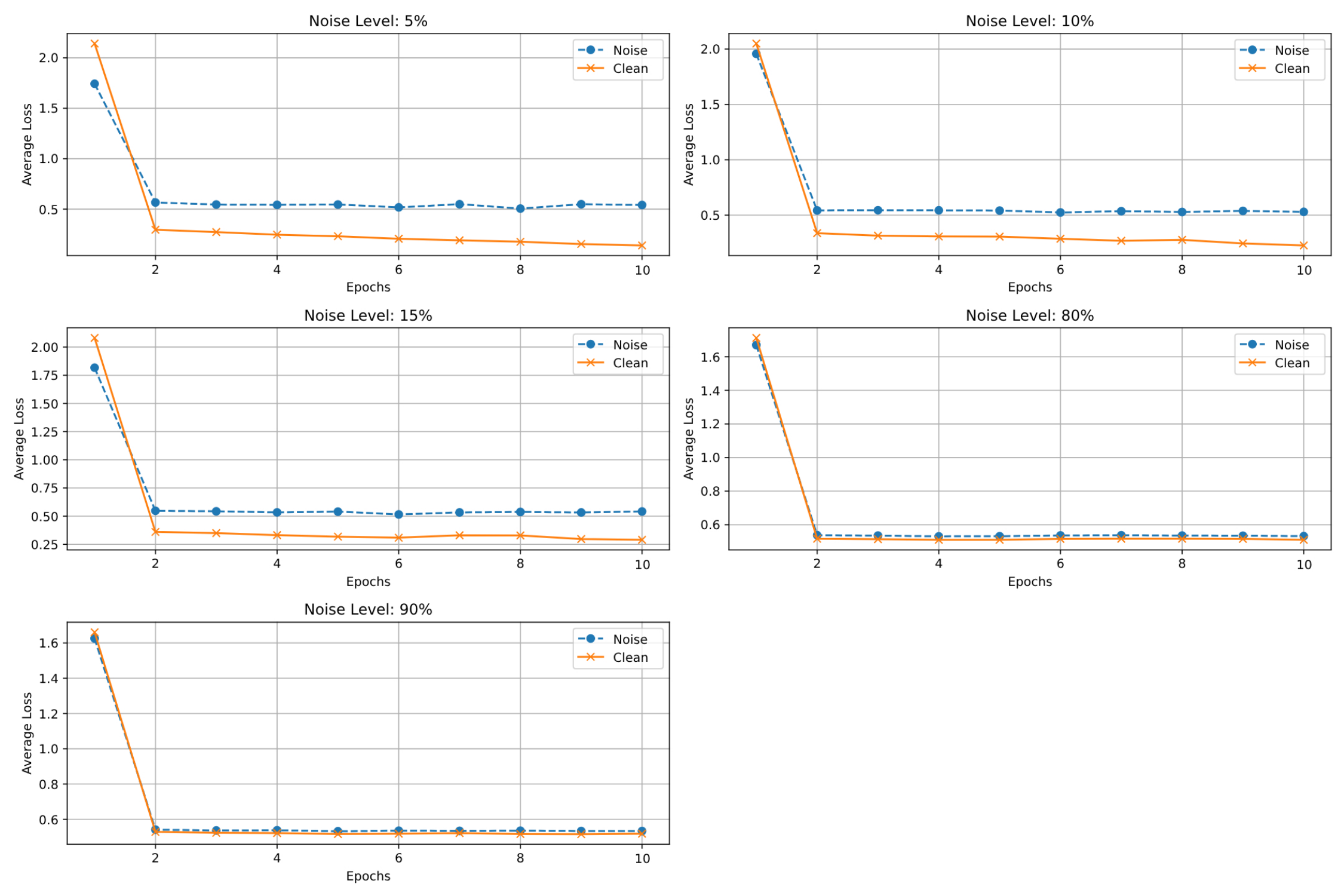} \caption{Loss Across Epochs (Subplots) for \ct} 
\label{fig:p_line} 
\end{figure}

In Figures~\ref{fig:r_line} and~\ref{fig:p_line} we show the difference between the loss of noise sample and clean sample with the discussed noise level, plus two more: 80\% and 90\% for \codebert~ and \ct, where the common feature of the training dynamics comparing the noise and clean samples can be seen. At the end of the first few epochs, the model quickly pulls down the average loss of most of the clean samples to a very low level, while processing the noise samples slowly, resulting in a relatively flat region where the loss for noise is stable. Taking 5\% and 10\% noise as an example, both models show that after the second round, the loss of clean samples drops to a very low value, and continues to drop slightly, while the noise samples stay at a high value for several epochs, and also continue to drop slightly, 
as seen in Figures~\ref{fig:r_line} and~\ref{fig:p_line}. This observation is important because it shows that the model will try to learn from the noise, so the more we train on the noise, the more it damages the performance.
This `fast-then-slow' separation pattern is the resonance effect of the conflict between the noise labels and the true distribution. The model quickly fits the dominant pattern of the clean data on the one hand, and is dragged down by the false signal of the noise samples on the other. Despite the numerical differences, \ct, with its architecture and parameters, converges on the majority of the samples in the second round, whereas \codebert~requires more epochs to see improvement on the clean samples. Still, the phenomenon of delayed convergence of the noisy samples is shown in both models. A higher proportion of noise raises the noise curve overall. As training progresses, the gap between the two noise levels at the final plateau narrows, suggesting that regardless of the proportion of noise, the model attempts to agree on these samples in a compromise loss region if training is sufficiently deep. However, this plateau is much higher than the minimum loss of the clean samples. This leads to varying degrees of decline in the model's performance for the test set.

This cross-model consistency reveals an important conclusion: \textit{label noise has a constant effect on the convergence of large-scale Transformer models.} Not only does it cause separation in the early stages of training, but it also constantly misleads the model away from the correct pattern. While there are slight differences across model, this separation in noise and clean data trajectory cuts across model types, highlighting the impact of noise on training stability and final performance across tasks and models.

\begin{figure}[t]
  \centering
  \includegraphics[width=\columnwidth]{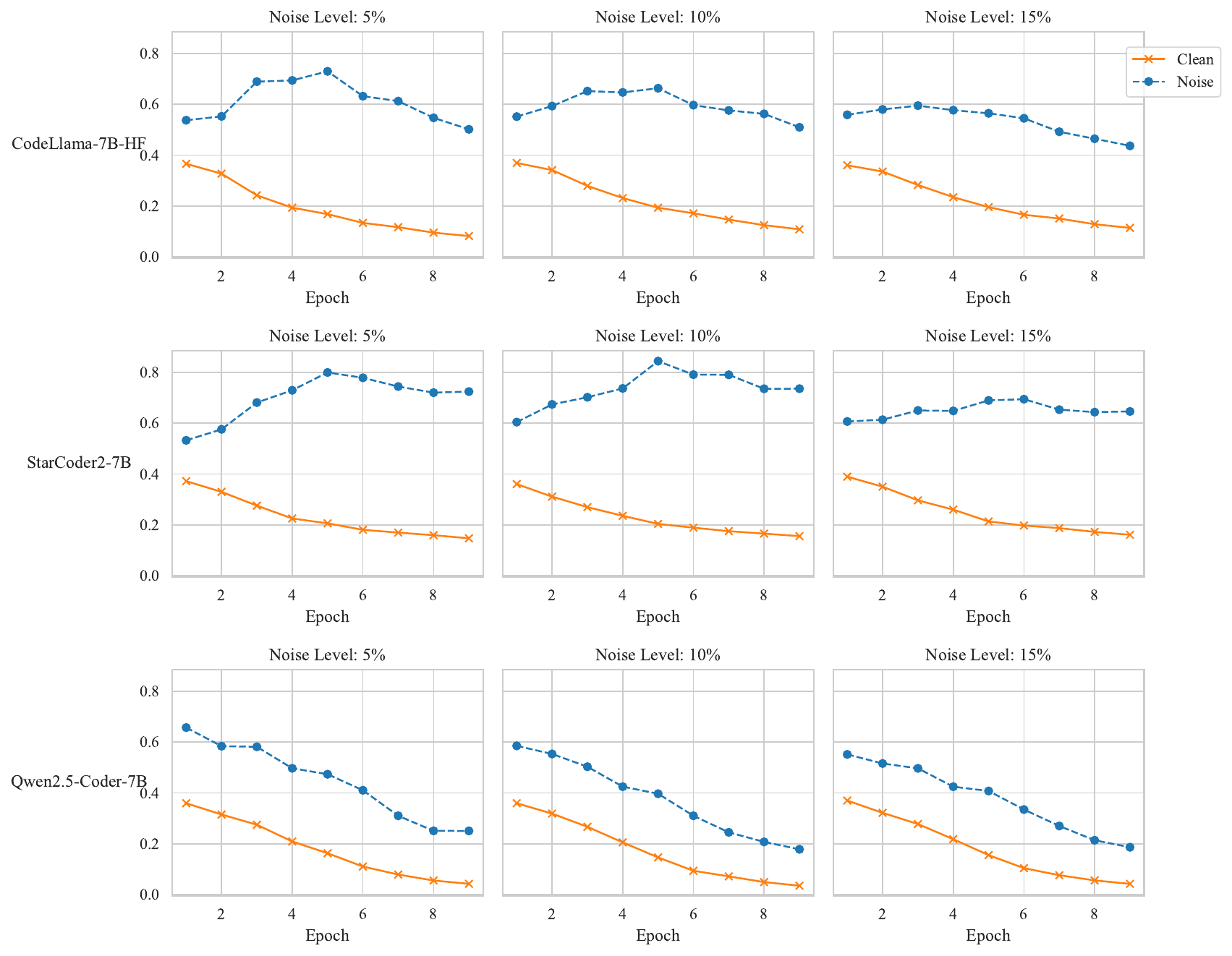}
  \caption{Average loss of clean versus noisy samples over epochs for three LLMs for commit intent classification at noise levels of 5\%, 10\%, and 15\%. Each row corresponds to one model and each column to one noise level; the solid (×) curve tracks clean‐sample loss and the dashed (o) curve tracks noisy‐sample loss.}
  \label{fig:noise-clean-loss-3x3}
  \vspace{-0.5em}
\end{figure}

When we look at LLMs for the code commit intention classification, another interesting point shows. Figure~\ref{fig:noise-clean-loss-3x3} shows the average loss of clean versus noisy samples over nine epochs for three LLMs for commit intent classification at noise levels of 5\%, 10\%, and 15\%. In the case of \codellama~(top row), we can see that the clean sample loss decays smoothly over all epochs, regardless of noise level, whereas the noisy sample loss behaves quite differently. 
At 5\% and 10\% noise, the noisy‐loss curve dips slightly, peaks around epoch 5, then declines, signaling that the model initially rejects corrupted labels by amplifying their loss before learning to separate them. At 15\% noise, that peak vanishes, and the noisy loss simply falls in lockstep with the clean loss. This implies that the sheer volume of spurious labels overwhelms the model’s ability to isolate them. Similar trend can be found in \starcoder, we see a similar two‐phase trajectory but with an even more pronounced peak in noisy sample loss between epochs two and five. In particular, at ten percent noise, the model amplifies error on the corrupted examples to a greater extent than \codellama, suggesting a stronger internal mechanism for flagging anomalous data. Yet once the noise ratio increases to fifteen percent, the height of that peak diminishes, suggesting that as noise levels rise beyond certain thresholds, models may struggle to maintain a clear distinction between clean and corrupted samples during fine-tuning.

In contrast, \qwen~(bottom row) shows no such transient resistance. Its noisy sample loss plummets from the first epoch and remains on a steady downward trend, closely tracking the clean sample loss, albeit at a consistently higher level. Even as noise increases, there is no pronounced rise or plateau, no evidence of the model disagreeing with mislabeled examples. 
This is also reflected in its F1 score: pure LoRA falls to 44.8\% at 15\% noise, whereas with MANTRA it holds steady at 52.3\%, the drop being narrowed considerably, and the F1 score is higher than pure LoRA at all noise levels. The performance of \qwen~highlights the universal value of MANTRA, which shows that dynamic sample removing not only reduces the interference of mislabeling, but also improves the robustness of the model that is sensitive to noise.
% \FHF{Great, but, what does this differnce tell us about qwen model? You should bring some discussions that why the other LLMs see such difference, but qwen does not. is this related to architecture or pre-training, etc? is this good or bad?}\shawn{This difference suggests important insights into \qwen's characteristics. Specifically, the continuous and smooth reduction in loss could indicate that \qwen'ss pre-training procedure or architectural choices result in internal representations that are inherently robust, but not explicitly discriminative against noise. Unlike other models, \qwen may smoothly incorporate all training samples, including corrupted ones, into its parameter space without clear rejection boundaries. This lack of noise resistance can be viewed both positively and negatively: positively, because it demonstrates\qwen's capacity to integrate noisy data without drastic degradation, implying strong generalization capabilities; negatively, because it suggests a potential vulnerability to noise accumulation, which can degrade performance significantly at higher noise levels.}

Overall, label noise broadens the loss distribution and slows convergence, degrading performance as noise increases, whereas MANTRA flattens the performance curves and even lets \starcoder~reclaim its lead under certain noise conditions. Models’ initial noise resistance and their reactions to dynamic culling reflect differences in capacity, pre‑training data richness, and architecture. Future work can tailor culling thresholds and schedules per architecture to optimize performance across noisy scenarios.

\begin{figure}[t]
  \centering
  \includegraphics[width=\columnwidth]{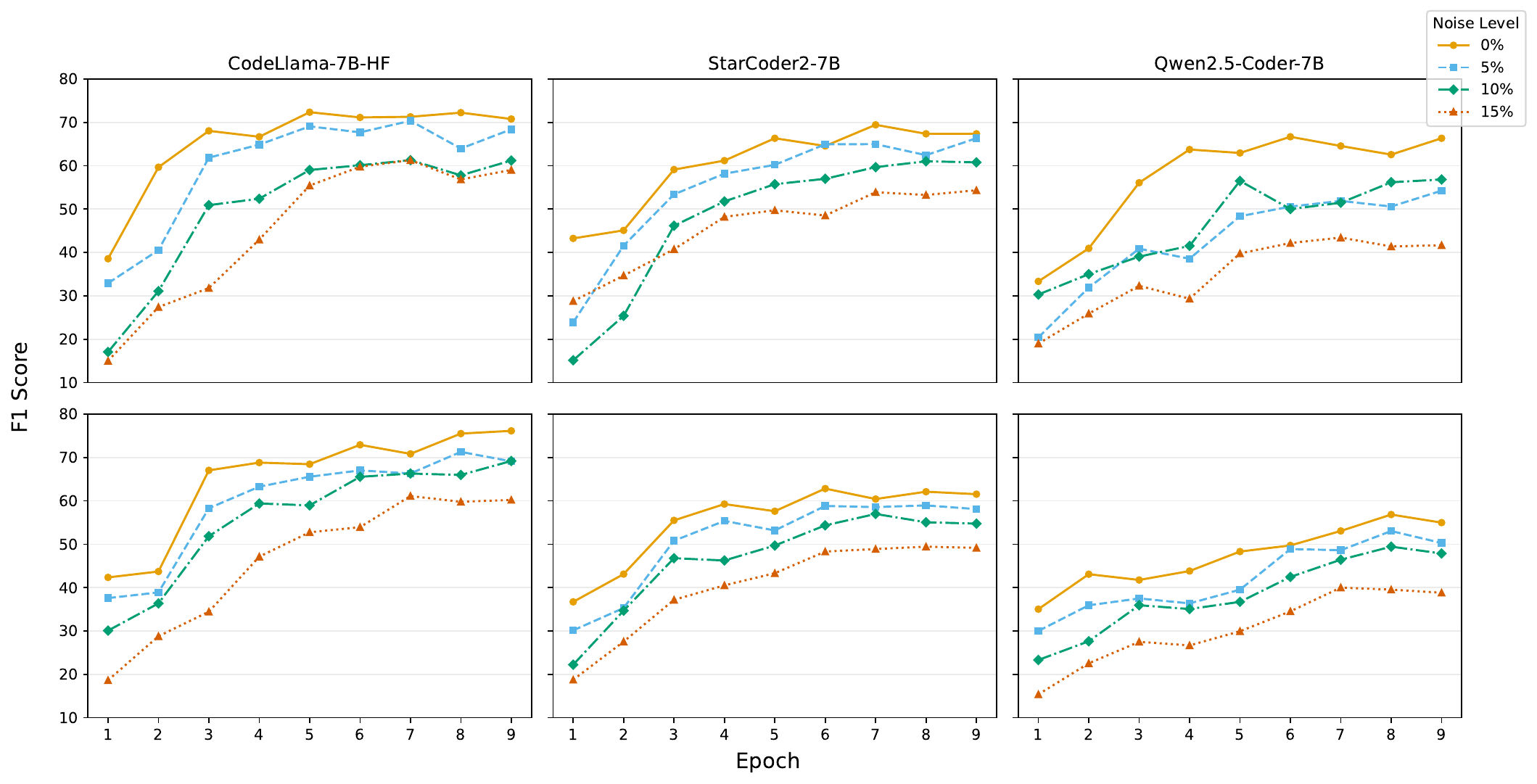}
  \caption{F1 score across epochs for three LLMs under four label-noise levels. The top row shows LoRA fine-tuning (“Without MANTRA”); the bottom row applies MANTRA.}
  \label{fig:f1_per_epoch}
  \vspace{-0.5em}
\end{figure}

Figure~\ref{fig:f1_per_epoch} shows F1 score across epochs for three LLMs under four label-noise levels. Across all LLMs, the F1 score on the validation set after each epoch without MANTRA (top row) exhibits clear signs of disruption when different levels of noise are inserted into the training set. In the clean setting (0\% noise), each model climbs steadily toward its peak F1 and stays stable when it reaches the peak. After injecting noise, however, the ascent breaks into a series of bumps and dips. At 5\% noise, the curves still eventually recover, but their peaks come later and they wobble more: \codellama's F1 falters around epoch 3 and again at epoch 8, \starcoder\ even briefly plateaus or dips under clean performance, and \qwen\'s 10\% noise curve spikes irregularly around epoch 5 before settling. As noise rises to 10\% and 15\%, these oscillations grow more pronounced and the maximum F1 shrinks. This behavior reflects the model oscillating between fitting the majority of clean examples and over‐memorizing the noise; it loses ground on true validation patterns when trying to fit to noise, then regains it when the clean signal briefly dominates.

By contrast, once MANTRA’s dynamic filtering is applied (bottom row), all three models’ F1 trajectories realign almost perfectly with the noise-free baseline. Even at 15\% noise, we now see a smooth upward trend that closely shadows the 0\% curve, with only a small, consistent gap in absolute F1. The jagged jumps disappear, the peaks return to the same epochs as the clean data, and the models regain confidence much earlier in training. In effect, MANTRA prevents the model from getting lost in corrupted labels by removing high‐loss examples so that each optimization step reinforces genuine patterns, not spurious ones.

Noisy labels induce high-loss examples that pull gradients in opposing directions, causing oscillations in validation loss and F1. By using a GMM to prune those outliers, MANTRA ensures each epoch’s updates mirror the clean-data distribution, yielding smooth, convergent F1 curves at any noise level.

\textbf{Comparing with existing works. }
% \FHF{Add the comparison with SE related works here. What you find that are similar to their findings? What aspects are different? What can we learn from your findings? What does it imply for researchers and practitioners?}
Both our work and~\citet{shah2024towards} observe that label noise leaves a clear impact in training dynamics: clean samples rapidly converge to low loss while noisy samples lag, producing broader tails or multiple loss distributions and delayed convergence.~\citet{shah2024towards}~report near‑zero biases in many layers and significant gradient instability when training on buggy data. In our loss density analysis, we similarly see a pronounced fast then slow separation, where clean losses plummet early, noisy losses plateau higher and much later, which underlies the instability~\citet{shah2024towards} document. Consistent with ~\citet{wang2024empirical}, we confirm that smaller models suffer severe performance degradation even under low noise rates (5–10\%). These PTMs lose accuracy quickly as label corruption increases. Whereas~\citet{wang2024empirical}~find that larger models remain largely robust to label noise, our experiments show that Qwen‑2.5‑Coder is notably sensitive. It's F1 on the validation set plummets sharply, even at moderate noise (10\%), and exhibits pronounced fluctuations across epochs. In contrast, CodeLlama and StarCoder2 still display the resilience ~\citet{wang2024empirical} describe. Our experiments show that embedding MANTRA during training produces smoother convergence and higher accuracy for different model sizes and architectures. For researchers, this underscores the value of dynamic, model‑aware noise mitigation strategies.

% \shawn{combines adaptive dropout with noise-resistant regularization techniques: we track each sample’s loss and fit a GMM to those per‑sample losses. One component corresponds to “clean” samples and the rest to “noisy” samples. Then, rather than apply a fixed dropout rate across all data, we use this clustering to adaptively drop out only the persistently high loss samples as training proceeds.}

% \shawn{Without compromising data diversity: Because the dropout is adaptive and delayed, we only begin excluding samples after the model has had an initial pass at everything, which means the model still sees the full variety of examples early on.}

%% file: sections/Threats.tex
\section{Threats to Validity}
\textbf{Internal Validity.}~Our experiments introduce noise by randomly flipping ground-truth labels uniformly. While this provides precise control and reproducibility, real-world annotation errors are typically non-uniform—developers show systematic biases, and automated pipelines introduce structured noise patterns. Our synthetic approach may not fully capture these real-world noise characteristics, though we mitigate this by testing multiple noise levels.

\textbf{External Validity.}
In our study, we examined two tasks of code summarization and commit intent classification. While our results might not generalize to other tasks, we chose two different types of tasks, one is a generative and another is classification. The variety of task types in our study alleviates the problems that MANTRA could only be applied to certain tasks. 
Another validity concern relates to the studied datasets. 
Other datasets and industrial code bases might have different data and noise distribution, deviating from the datasets we examined and the simulated noise. 
Though we selected widely-used benchmarks, enhancing comparability and relevance across research, the efficacy of MANTRA may need recalibration. 
That being said, our approach is not specific to a tasks or model, therefore, we anticipate low efforts of applying MANTRA on new tasks and datasets. 

\textbf{Hardware limitation} We have targeted five popular Transformer models up to 7 B parameters. However, due to hardware limitations, we have to use LoRA and quantization during fine-tuning, which freezes the vast majority of the pretrained weights. While LoRA has shown results in some experiments that are similar to or even exceed those of full fine-tuning~\cite{lora}, full-parameter fine-tuning could either amplify noise memorization, making MANTRA even more necessary, or conversely absorb noise more smoothly, reducing the relative benefit of our approach.

%% file: sections/conclusion.tex
\section{Conclusion}

In this study, we examined how controlled label noise at 0\%, 5\%, 10\%, and 15\% alters the behavior of five code models on code summarisation and commit‑intent classification. Noise broadened the loss landscape, postponed the convergence of corrupted examples, and reduced task accuracy. To counteract this effect we introduced MANTRA, a Gaussian‑mixture filter that operates during fine‑tuning, whether full or LoRA, and is paired with dropout. MANTRA restored a clear separation between clean and noisy samples: at the highest noise level, CodeLlama and CodeT5+ lost under three per cent BLEU or F1, StarCoder recovered its lead, and even the most fragile model, Qwen, regained more than half of its lost accuracy while dropout kept it from over‑fitting spurious patterns. The study shows that label noise invariably splits the learning dynamics of clean and corrupted data, that sequence‑to‑sequence architectures first resist and then assimilate noise, and that adaptive filtering with light regularisation can almost remove the performance penalty imposed by severe corruption.

Future work will refine mixture thresholds, explore alternative filters, adjust dropout schedules, and apply the method to naturally noisy corpora such as issue trackers and pull‑request discussions.